\documentclass[journal]{IEEEtran}

\usepackage{subfigure}
\usepackage{setspace}
\usepackage{multirow} 
\usepackage{algorithm}
\usepackage{algorithmic}
\usepackage{amsmath}
\usepackage{amssymb}
\usepackage{latexsym}
\usepackage{multirow}
\usepackage{epsfig}
\usepackage{graphics}
\usepackage{graphicx}
\usepackage{mathrsfs}
\usepackage{subfigure}
\usepackage{slashbox}
\usepackage{mathrsfs}
\usepackage{bbding}
\usepackage{cite}
\usepackage{color}
\usepackage{xcolor}
\usepackage{booktabs}
\usepackage{amssymb,mathrsfs,amsmath}
\newcommand{\bm}[1]{\mbox{\boldmath{$#1$}}}
\usepackage{flushend}
\usepackage{stfloats}
\usepackage{hyperref}
\usepackage{amsthm}
\theoremstyle{plain}
\newtheorem{rem}{Remark}

\allowdisplaybreaks [4]

\begin{document}	
	
\title{Joint Transmitter and Receiver Design for Movable Antenna Enhanced Multicast Communications}
\author{Ying~Gao, Qingqing~Wu,~\IEEEmembership{Senior Member,~IEEE}, and Wen~Chen,~\IEEEmembership{Senior Member,~IEEE}\vspace{-4mm} 
\thanks{The work of Qingqing Wu was supported by National Key R\&D Program of China (2022YFB2903500), NSFC 62371289, NSFC 62331022, and Xiaomi Young Scholar Program. The work of Wen Chen was supported by NSFC 62071296 and Shanghai Kewei 22JC1404000. \emph{(Corresponding author: Qingqing Wu.)} }
\thanks{The authors are with the Department of Electronic Engineering, Shanghai Jiao Tong University, Shanghai 201210, China (e-mail: yinggao@sjtu.edu.cn; qingqingwu@sjtu.edu.cn; whenchen@sjtu.edu.cn).}}

\maketitle

\begin{abstract}
	Movable antenna (MA) is an emerging technology that utilizes localized antenna movement to achieve better channel conditions for enhancing communication performance. In this paper, we study the MA-enhanced multicast transmission from a base station equipped with multiple MAs to multiple groups of single-MA users. Our goal is to maximize the minimum weighted signal-to-interference-plus-noise ratio (SINR) among all the users by jointly optimizing the position of each transmit/receive MA and the transmit beamforming. To tackle this challenging problem, we first consider the single-group scenario and propose an efficient algorithm based on the techniques of alternating optimization and successive convex approximation. Particularly, when optimizing transmit or receive MA positions, we construct a concave lower bound for the signal-to-noise ratio (SNR) of each user using only the second-order Taylor expansion, which simplifies the problem-solving process compared to the existing two-step approximation method. The proposed design is then extended to the general multi-group scenario. Simulation results show that the proposed algorithm converges faster than the existing two-step approximation method, achieving a 3.4\% enhancement in max-min SNR. Moreover, it can improve the max-min SNR/SINR by up to 22.5\%, 181.7\%, and 343.9\% compared to benchmarks employing only receive MAs, only transmit MAs, and both transmit and receive FPAs, respectively. 
\end{abstract}


\section{Introduction}

In recent decades, wireless communication has evolved from single-input-single-output (SISO) to multiple-input multiple-output (MIMO) systems \cite{2017_Busari_MIMO}. By utilizing independent or quasi-independent channel fading from multipath components, MIMO systems enable simultaneous transmission of multiple data streams within the same time-frequency resource block, significantly enhancing spectral efficiency compared to single-antenna systems \cite{1999_Telatar_MIMO,2004_Paulraj_MIMO,2003_Goldsmith_MIMO}. However, conventional MIMO systems with fixed-position antennas (FPAs) face limitations in diversity and spatial multiplexing performance due to their static and discrete arrangement, restricting full exploitation of channel variations in the continuous spatial field. More recently, the wireless communication research community has devoted significant attention to two antenna technologies: ``fluid antenna'' (FA) and ``movable antenna'' (MA). This interest is driven by their unique flexibility and reconfigurability, which offer significant improvements in system performance for wireless applications. In fact, the term ``fluid antenna'' first emerged in a 2008 US patent \cite{2008_Tam_FLA}, while the term ``movable antenna'' was introduced in a book \cite[Section 17.4]{2008_Pan_MA} published in the same year. Originally, FA referred to antennas utilizing fluid dielectrics as electromagnetic radiators, while MA referred to antennas endowed with motion/rotation capability through micro-electromechanical systems. Despite their longstanding presence in antenna technology, it is only in recent years that they have been studied systematically within the field of wireless communication research \cite{2021_Wong_fluid,2020_Wong_Fluid,2023_Lipeng_Modeling,2023_Wenyan_MIMO,2023_Lipeng_overview}. Moreover, the concepts of FA and MA have evolved beyond their original definitions to encompass all types of antennas that can be flexibly adjusted in a spatial region. Although FA and MA stem from distinct origins and may be realized differently in practice, they share the same mathematical model of flexible antenna positioning and have been treated as interchangeable terms in scholarly works. 

In this paper, we adopt the term MA. In MA systems, each antenna is connected to the radio frequency chain via a flexible cable, and its position can be flexibly adjusted within a spatial region with the aid of mechanical controllers and drivers \cite{2023_Lipeng_overview}. Thus, unlike FPAs, which remain static and undergo random wireless channel variations, MAs can be strategically deployed at positions with better channel conditions to improve communication performance. In particular, MAs offer superior signal power enhancement, interference mitigation, flexible beamforming, and spatial multiplexing capabilities compared to FPAs \cite{2023_Lipeng_uplink}.

Additionally, compared with the conventional antenna selection (AS) technique, which can also achieve spatial diversity gains, MAs are more cost-effective and efficient for two main reasons. Firstly, AS requires a large number of candidate antennas to select from, whereas MAs can fully utilize the spatial diversity in a given region with significantly fewer antennas. Secondly, in AS systems, antennas are typically confined to discrete deployment at fixed locations, either arranged in a one-dimensional (1D) line or a two-dimensional (2D) surface. In contrast, MAs are capable of moving freely within a three-dimensional (3D) space, offering greater flexibility and improved performance  \cite{2023_Lipeng_overview,2004_Molisch_AS,2004_Sanayei_AS}.

The attractive advantages of MAs have generated significant interest in integrating them into various systems,  particularly for optimizing their positions to achieve different design objectives. However, existing research has predominantly concentrated on unicast transmissions. In real-world wireless networks, groups of users often require identical data in various scenarios, such as live sports broadcasts, mobile TV, system updates, and popular video streaming \cite{2012_David_multicast}. Relying solely on unicast would lead to the same data being sent multiple times, thereby decreasing spectrum and energy efficiency. By contrast, multicast can deliver the same data to multiple users simultaneously, making it more efficient for serving users with shared interests. Despite this, the potential of MAs for enhancing multicast communications has yet to be fully explored. \looseness=-1

\subsection{Related Works}\label{Sec_intro_realted}
In this subsection, we review existing works from two perspectives: MA-enabled systems and multicast systems. 
\subsubsection{MA-Enabled Systems} 
Initial studies in this field began with point-to-point single-user MA-enabled systems. For example, the authors of \cite{2023_Lipeng_Modeling} proposed the mechanical MA architecture and a field-response for a SISO system consisting of a single-MA transmitter and a single-MA receiver. They also analyzed the maximum signal-to-noise ratio (SNR) gain achieved by the single receive MA over its FPA counterpart under both deterministic and stochastic channels, assuming that the transmit MA is fixed at the reference point. In \cite{2023_Wenyan_MIMO}, the authors investigated an MA-enabled point-to-point MIMO system. Their numerical results showed that MA-enabled MIMO systems can achieve significantly higher capacity over conventional FPA-based MIMO systems with/without AS. Then, the works in \cite{2023_Xintai_statistical} and \cite{2024_Yuqi_Fluid_statistical} expanded upon the research conducted in \cite{2023_Wenyan_MIMO} to address scenarios where only statistical channel state information (CSI) is available.  Besides, for MA-enabled single-user systems, researchers investigated scenarios with one eavesdropper in \cite{2024_Zhenqiao_secure,2024_Jun_MAsecure}, and with multiple eavesdroppers in  \cite{2024_Guojie_secure_lett,2024_Guojie_secure_long}. Unlike the aforementioned narrow-band studies, the authors of \cite{2024_Lipeng_wideband} explored MA-aided wideband SISO communications.

Meanwhile, there have been some studies on more general MA-assisted multiuser systems, focusing on uplink scenarios such as those in \cite{2023_Lipeng_uplink, 2023_Guojie_gradient,2023_Zhenyu_uplink,2024_Nian_MA}. For instance, the authors of \cite{2023_Lipeng_uplink} examined the MA-enhanced uplink transmission from multiple single-MA users to a base station (BS) equipped with an FPA array, demonstrating significant reductions in total transmit power compared to FPA systems. Beyond uplink scenarios, works such as \cite{2023_Yunan_RIS_MA,2023_Guojie_CoMP,2024_Zhenqiao_Fluid,2024_Haoran_MA,2024_Songjie_multisuer,2024_Caihao_Learning,2024_Weidong_graph,2024_Yichi_hybrid,2023_Yifei_discrete} have explored MA-enabled multiuser downlink systems. Moreover, the authors of \cite{2024_Bin_symbiotic} and \cite{2024_Jinze_full-duplex} studied MA-enabled symbiotic radio and full-duplex communications, respectively. In addition to the above works that are limited to single-cell scenarios, the authors of \cite{2024_Honghao_IFC} investigated an MA-aided multi-cell interference channel. On the other hand, some research has explored channel estimation for MA-enabled communications (see, e.g., \cite{2023_Wenyan_estimation,2023_Zhenyu_estimation,2023_Zijian_estimation}).

\subsubsection{Multicast Systems}
In the seminal study on multicasting \cite{2006_Nikos_multicast}, the authors investigated precoder design for max-min fairness (MMF) and quality of service (QoS) problems in a single-group system. They demonstrated that both problems are NP-hard and proposed suboptimal solutions using semidefinite relaxation combined with randomization. This work was extended to multi-group systems in \cite{2008_Eleftherios_multicast}, where a duality between the MMF and QoS problems was discovered. Further, the authors of \cite{2014_Dimitrios_multicast} established a similar duality under the per-antenna peak power constraints, making the model more practical. Moreover, alternative solutions based on convex approximation methods for semidefinite relaxation were proposed in \cite{2011_Nilst_multicast,2012_Schad_multicast}. These solutions exhibit marginal performance improvements in certain scenarios while maintaining lower complexities. Also, multicast beamforming problems were studied in various scenarios, including multi-cell coordination networks \cite{2013_Zhengzheng_multicast}, relay networks \cite{2011_Bornhorst_multicast}, cache-enabled cloud radio access network \cite{2016_Tao_multicast}, etc.

It is worth noting that in a multicast system, the data rate of each group is constrained by the user with the worst channel condition. Therefore, if any link encounters a poor channel, the overall multicast capacity may be greatly reduced. In recent years, intelligent reflective surface (IRS) \cite{2019_Qingqing_Joint} has been considered as a promising technology to improve the performance of multicast systems. An IRS consists of numerous reflecting metamaterial elements. By intelligently adjusting the reflection coefficient of each element according to the changing environment, the IRS can improve the reception of desired signals and reduce interference for the receiver \cite{2020_Qingqing_IRS_Intro}. The authors of \cite{2020_Gui_multicast} studied the sum rate maximization problem for an IRS-aided multicast system. Their simulation results showed that incorporating the IRS can significantly improve both spectral and energy efficiency. Additionally, the authors of \cite{2022_Liangsen_multicast} and \cite{2022_Linsong_multicast} extended this research by exploring IRS-aided multicast systems in more complex scenarios, such as in the presence of multiple eavesdroppers and an interferer, respectively.

\subsection{Motivation and Contributions}
Despite several works demonstrating the advantages of employing MAs in various systems, research on MA-enhanced multicast systems is still in its early stages. The spatial diversity and interference mitigation gains provided by MAs are expected to enable MA-aided multicast systems to significantly outperform their FPA counterparts. Although both IRS and MA have the potential to improve multicast system performance, there are notable differences between them. Specifically, IRS reconfigures the propagation environment by reflecting incident signals in preferred directions. In other words, IRS does not have the capability to transmit signals on its own and cannot function independently of the communication system. Integrating IRS into communication systems introduces an additional device, which increases system complexity and potentially affects overall reliability. By contrast, MA creates favorable channel conditions through local antenna movement using mechanical controllers and drivers, without requiring the introduction and maintenance of third-party devices. However, MA's flexibility is limited: it can only move within a finite region at the transmitter/receiver, while IRS can be placed at any location. Overall, MA and IRS are complementary technologies to each other. Our work on MA-enhanced multicast systems can pave the way for exploring the synergy between MA and IRS in multicast systems.

On the other hand, among the 28 works related to MAs mentioned in the previous subsection, aside from the three studies on channel estimation, most focus only on transmit MA(s) with fixed receive FPA(s) (or fixed transmit FPA(s) with receive MA(s)). Only a few, such as \cite{2023_Wenyan_MIMO}, \cite{2023_Xintai_statistical}, \cite{2024_Yuqi_Fluid_statistical}, and \cite{2024_Lipeng_wideband}, consider the joint optimization of transmit MAs and receive MAs. However, even these studies do not determine whether employing only transmit MAs or employing only receive MAs yields better performance. 
The comparison results can provide important engineering insights. For example, if implementation complexity and hardware costs restrict the choice to employing either transmit MAs or receive MAs, but not both, then understanding which option delivers better performance can guide decision-making. Theoretically speaking, for the case employing receive MAs, it is feasible for all users to achieve their individual maximum SNR/signal-to-interference-plus-noise ratio (SINR) simultaneously through strategic positioning of receive MAs. In contrast, for the case employing transmit MAs, users are consistently in competition for resources, and there is always a non-trivial trade-off among users' SNR/SINR. More specifically, the positioning of each transmit MA needs to weigh the trade-offs among the channel conditions of all users, and improving the SNR/SINR of one user often comes at the cost of decreasing the SNR/SINR of others. Thus, receive MAs may provide a more significant performance enhancement in achievable max-min SNR/SINR than transmit MAs, even when their number is slightly less than that of transmit MAs, let alone when their number exceeds that of transmit MAs.

\begin{figure*}[!t]
	\centering
	\includegraphics[scale=0.66]{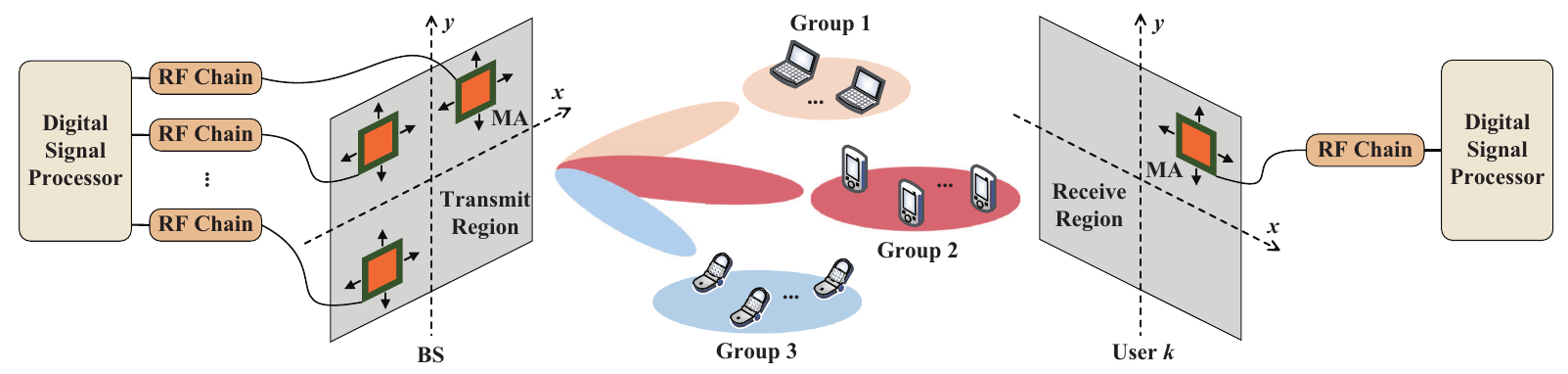}
	\caption{Illustration of an MA-enhanced multi-group multicast MISO communication system.} \label{Fig:system_model}
\end{figure*} 

Motivated by the above discussions, this paper investigates an MA-enabled multicast multiple-input single-output
(MISO) communication system which is composed of a BS with $M$ transmit MAs and $N$ groups with a total of $K$ single-MA users overall, as shown in Fig. \ref{Fig:system_model}. The main contributions of this paper are summarized as follows: 
\begin{itemize}
	\item To the best of the authors' knowledge, this is the first work in the literature to study the joint transmitter and receiver design for MA-enhanced multicast transmission. We formulate a minimum weighted SINR maximization problem that jointly optimizes the positions of transmit/receive MAs and the transmit beamforming, under the constraints of maximum transmit power at the BS, finite moving region for each MA, and minimum inter-MA distance. 
	\item To handle the non-convex and challenging optimization problem, we first consider a simplified scenario with only one group and propose an efficient AO algorithm, which will serve as a building block for the general multi-group scenario. Specifically, the transmit precoder, the position of each transmit MA, and the positions of all the receive MAs are iteratively optimized based on the successive convex approximation (SCA) technique. In particular, when optimizing transmit or receive MA positions, we construct a concave lower bound for each user's SNR by applying only the second-order Taylor expansion. This distinguishes our proposed algorithm from the two-step approximation method in \cite{2023_Wenyan_MIMO} (and its follow-up works such as \cite{2024_Yuqi_Fluid_statistical} and \cite{2024_Nian_MA}), which requires applying the first-order Taylor expansion before the second-order Taylor expansion. Subsequently, we extend the proposed algorithm to the general case with multiple groups by introducing slack variables.  
	\item Numerical results demonstrate that: 1) Our proposed algorithm converges more rapidly than the method presented in \cite{2023_Wenyan_MIMO} and achieves a 3.4\% improvement in max-min SNR. 2) In comparison to benchmarks employing only receive MAs, only transmit MAs, and both transmit and receive FPAs, the proposed algorithm can boost the max-min SNR/SINR by up to 22.5\%, 181.7\%, and 343.9\%, respectively. 3) There is a notable reduction in the required amount of transmit power or antennas to achieve a target level of max-min SNR/SINR performance with the proposed algorithm compared to benchmark schemes. 4) In overloaded scenarios where $M < N$, employing only transmit MAs typically yields marginal improvements in max-min SINR compared to using both transmit and receive FPAs. In contrast, employing only receive MAs can still provide notable performance gains, such as improving max-min SINR from $-0.29$ dB to $3.33$ dB. 5) Under the assumption that the transmit region and each receive region have identical sizes, employing only receive MAs results in higher max-min SNR/SINR than employing only transmit MAs when $M \leq K$, and this remains true even when $M$ is slightly larger than $K$. This supports the previous theoretical predictions about the performance comparison between employing only receive MAs and only transmit MAs.
\end{itemize}

\subsection{Organization and Notations}
The rest of this paper is organized as follows. Section \ref{Sec:model_and_formulation} introduces the system model and formulates the minimum SINR maximization problem for an MA-enabled multi-group multicast communication system. An efficient algorithm is proposed in Section \ref{Sec:Single} for addressing the simplified scenario with only one group, which is subsequently extended to handle the general multi-group scenario in Section \ref{Sec:Multi}. We evaluate
the performance of our proposed algorithm via simulations in Section \ref{Sec:Simulation}. Finally, Section \ref{Sec:Conclusion} concludes the paper.  

\emph{Notations:} $\mathbb C$ denotes the complex space. $\mathbb C^{M\times N}$ represents the space of $M\times N$ complex-valued matrices. For a complex-valued number $x$, $\left|x\right|$, ${\rm Re}\{\cdot\}$, and $\arg\left(x\right) $ denote its modulus, real part, and phase argument, respectively. For a complex-valued vector $\mathbf a$, $\left\|\mathbf a \right\|$ and $\left[\mathbf a\right]_i$ denote its Euclidean norm and $i$-th element, respectively. For a matrix $\mathbf A$ of arbitrary size, $\left\|\mathbf A\right\|_2$, $\left\|\mathbf A\right\|_F$, and $\left[\mathbf A\right]_{i,j}$ represent its spectral norm, Frobenius norm, and $(i,j)$-th element, respectively. For two square matrices $\mathbf S_1$ and $\mathbf S_2$, $\mathbf S_1 \succeq \mathbf S_2$ indicates that $\mathbf S_1 - \mathbf S_2$ is positive semidefinite. For a set $\mathcal X$, $\left|\mathcal X\right|$ denotes its cardinality. $\mathbf I$ is an identity matrix whose dimension is determined by the context. The conjugate transpose operator is denoted by $(\cdot)^H$, while the expectation operator is represented by $\mathbb E\left(\cdot\right)$. ${\rm diag}\left(\cdot\right)$ denotes the diagonalization operation. $\mathcal{CN}\left(\mathbf x,\mathbf \Sigma\right)$ represents a complex Gaussian distribution with a mean vector $\mathbf x$ and co-variance matrix $\mathbf \Sigma$. $\jmath \triangleq \sqrt{-1}$ refers to the imaginary unit. 

\section{System Model and Problem Formulation}\label{Sec:model_and_formulation}
\subsection{System Model}
As shown in Fig. \ref{Fig:system_model}, we consider an MA-enabled multi-group multicast communication system, where a BS equipped with $M$ transmit MAs serves $K$ single-MA users grouped into $N$ ($1 \leq N \leq K$) multicast groups. The set of transmit MAs and users are denoted by $\mathcal M$ and $\mathcal K$, respectively, with $\left|\mathcal M\right| = M$ and $\left|\mathcal K\right| = K$. Let $\mathcal G_n$ denote the set of users in group $n$, $n\in\mathcal N\triangleq \{1,\cdots, N\}$. Each user belongs to only one group, hence $\mathcal G_n \cap \mathcal G_q = \emptyset, \forall n,q\in\mathcal N$ and $\bigcup_{n=1}^N\mathcal G_n = \mathcal K$. In addition, the transmit and receive MAs are connected to RF chains via flexible cables, enabling them to move freely within local regions.\footnote{For the hardware architecture of MA-mounted transmitters/receivers, interested readers can refer to \cite[Fig. 2]{2023_Lipeng_overview}. Aligning with existing works \cite{2023_Lipeng_Modeling} and \cite{2023_Wenyan_MIMO}, we assume that the positions of MAs can be adjusted freely and perfectly within given continuous regions to quantify the performance limit. This assumption establishes an upper bound for the performance of practical systems involving the discrete movement limitations of stepper motors, and the insights gained can guide the system design.} 
Let $\mathcal C^{\rm t}$ denote the given 2D moving region for the $M$ transmit MAs. The 2D moving region for the single MA at user $k$ is denoted by $\mathcal C^{\rm r}_k$. Without loss of generality, we assume that both $\mathcal C^{\rm t}$ and $\mathcal C^{\rm r}_k$ are square regions of size $A \times A$. The positions of the $m$-th transmit MA at the BS and the single receive MA at the $k$-th user are represented by $\bm t_m = \left[x_m^{\rm t}, y_m^{\rm t}\right]^T\in\mathcal C^{\rm t}$ and $\bm r_k = \left[x_k^{\rm r}, y_k^{\rm r}\right]^T 
\in\mathcal C^{\rm r}_k$, respectively. The reference points in the regions $\mathcal C^{\rm t}$ and $\mathcal C_k^{\rm r}$ are denoted by $\bm o^{\rm t} = [0,0]^T$ and $\bm o_k^{\rm r} = [0,0]^T$, respectively. 

The assumption is made that the sizes of the antenna moving regions are considerably smaller than the signal propagation distance, ensuring that the far-field condition is satisfied between the BS and the users. In this case, for each channel path component, altering the MA positions has no impact on the angle of departure (AoD), the angle of arrival (AoA), and the amplitude of the complex coefficient, but solely influences the phase of the complex coefficient. Denote the total number of transmit and receive channel paths from the BS to user $k$ as $L_k^{\rm t}$ and $L_k^{\rm r}$, respectively. The elevation and azimuth AoDs of the $i$-th transmit path for user $k$ are denoted by $\theta_{k,i}^{\rm t} \in \left[-\frac{\pi}{2}, \frac{\pi}{2}\right] $ and $\phi_{k,i}^{\rm t} \in \left[-\frac{\pi}{2}, \frac{\pi}{2}\right]$, respectively. The elevation and azimuth AoAs of the $j$-th receive path for user $k$ are denoted by $\theta_{k,j}^{\rm r} \in \left[-\frac{\pi}{2}, \frac{\pi}{2}\right]$ and $\phi_{k,j}^{\rm r} \in \left[-\frac{\pi}{2}, \frac{\pi}{2}\right]$, respectively. Then, for user $k$, the propagation distance difference of the $i$-th transmit path between positions $\bm t_m$ and $\bm o^{\rm t}$ is given by $\bm t_m^T\left[\cos\theta_{k,i}^{\rm t}\sin\phi_{k,i}^{\rm t}, \sin\theta_{k,i}^{\rm t}\right]^T \triangleq \bm t_m^T\bm a_{k,i}^{\rm t}$ \cite{2023_Lipeng_Modeling}. Accordingly, the phase difference of the $m$-th transmit path between positions $\bm t_m$ and $\bm o^{\rm t}$ is obtained as $\frac{2\pi}{\lambda}\boldsymbol t_m^T\boldsymbol a_{k,i}^{\rm t}$, where $\lambda$ denotes the carrier wavelength. To account for the phase differences in all $L_k^{\rm t}$ transmit paths from the BS to user $k$, the field-response vector of the $m$-th MA at the BS is written as 
\begin{align}\label{eqv:t_FRV}
	\bm g_k(\bm t_m) = \left[e^{\jmath\frac{2\pi}{\lambda}\boldsymbol t_m^T\boldsymbol a_{k,1}^{\rm t}}, e^{\jmath\frac{2\pi}{\lambda}\boldsymbol t_m^T\boldsymbol a_{k,2}^{\rm t}}, \cdots, e^{\jmath\frac{2\pi}{\lambda}\boldsymbol t_m^T\boldsymbol a_{k,L_k^{\rm t}}^{\rm t}}\right]^T.
\end{align}
We stack $\left\lbrace\bm g_k(\bm t_m)\right\rbrace$ into a matrix to obtain the field-response matrix at the BS, represented as
\begin{align}
	\bm G_k(\bm t) = \left[\bm g_k(\bm t_1), \bm g_k(\bm t_2), \cdots, \bm g_k(\bm t_M)\right] \in \mathbb C^{L_{k}^{\rm t}\times M},
\end{align}
where $\bm t \triangleq \{\bm t_m\}$. Similar to \eqref{eqv:t_FRV}, the field-response vector of the receive MA at the $k$-th user is given by 
\begin{align}
	\bm f_k(\bm r_k) = \left[e^{\jmath\frac{2\pi}{\lambda}\boldsymbol r_k^T\boldsymbol a_{k,1}^{\rm r}}, e^{\jmath\frac{2\pi}{\lambda}\boldsymbol r_k^T\boldsymbol a_{k,2}^{\rm r}}, \cdots, e^{\jmath\frac{2\pi}{\lambda}\boldsymbol r_k^T\boldsymbol a_{k,L_k^{\rm r}}^{\rm r}}\right]^T,
\end{align}
where $\bm a_{k,j}^{\rm r} \triangleq \left[\cos\theta_{k,j}^{\rm r}\sin\phi_{k,j}^{\rm r}, \sin\theta_{k,j}^{\rm r}\right]^T$. Furthermore, let $\mathbf \Sigma_k \in \mathcal C^{L_k^{\rm r}\times L_k^{\rm t}}$ represent the path-response matrix, which characterizes the responses between all the transmit and receive channel paths from $\bm o^{\rm t}$ to $\bm o_k^{\rm r}$. Then, the channel vector between the BS and the $k$-th user can be expressed as 
\begin{align}
	\bm h_k(\bm t, \bm r_k)^H = \bm f_k(\bm r_k)^H\mathbf \Sigma_k\bm G_k(\bm t) \in\mathbb C^{1\times M}. 
\end{align}

For multi-group multicast systems, users within the same group share the same message, while the messages required by different groups are independent of each other. At the BS, we consider a linear transmit precoding with $\bm w_n$ denoting the beamforming vector for group $n$. The transmitted signal from the BS is then given by $\bm x = \sum_{n=1}^N \bm w_ns_n$, where $s_n$ denotes the data symbol for group $n$ and $\{s_n\}$ are independent over $n$, satisfying $\mathbb E\left(\left|s_n\right|^2\right) = 1$, $\forall n\in\mathcal N$. The corresponding received signal at user $k\in\mathcal G_n$ can be written as
\begin{align}
	\hspace{-1mm} y_k = \bm h_k(\bm t, \bm r_k)^H\bm w_ns_n + \sum_{q=1,q\neq n}^N\bm h_k(\bm t, \bm r_k)^H\bm w_qs_q + z_k, 
\end{align}
where $z_k \sim \mathcal{CN}\left(0,\sigma_k^2\right)$ denotes the additive white Gaussian noise with variance $\sigma_k^2$ at user $k$. As a result, the SINR of user $k\in\mathcal G_n$ is given by
\begin{align}
	\text{SINR}_k = \frac{\left|\bm h_k(\bm t, \bm r_k)^H\bm w_n\right|^2}{\sum_{q=1,q\neq n}^N\left|\bm h_k(\bm t, \bm r_k)^H\bm w_q\right|^2 + \sigma_k^2}.
\end{align}

\subsection{Problem Formulation}
In this paper, we aim to maximize the minimum weighted SINR among all the users by jointly optimizing the transmit precoders $\{\bm w_n\}$, the transmit MA positions $\{\bm t_m\}$, and the receive MA positions $\{\bm r_k\}$. The problem of interest can be formulated as
\begin{subequations}\label{P1}
	\begin{eqnarray}
		\hspace{-2mm}\text{(P1)}: \hspace{-3mm}&\underset{\substack{\eta, \{\boldsymbol w_n\},\\ \{\boldsymbol t_m\}, \{\boldsymbol r_k\}}}{\max}&  \eta\\
		&\text{s.t.}& \hspace{-5mm}\frac{1}{\gamma_k}\frac{\left|\bm h_k(\bm t, \bm r_k)^H\bm w_n\right|^2}{\sum_{q=1,q\neq n}^N\left|\bm h_k(\bm t, \bm r_k)^H\bm w_q\right|^2 + \sigma_k^2} \geq \eta, \nonumber\\
		&& \hspace{-5mm} \forall k\in\mathcal G_n, n\in\mathcal N, \label{P1_cons:b}\\
		&& \hspace{-5mm} \sum_{n=1}^N\left\|\bm w_n\right\|^2 \leq P_{\max}, \label{P1_cons:c}\\
		&& \hspace{-5mm} \bm t_m \in\mathcal C^{\rm t}, \ \forall m\in\mathcal M, \label{P1_cons:d}\\
		&& \hspace{-5mm} \left\|\bm t_m - \bm t_p\right\| \geq D, \ \forall m,p\in\mathcal M, m\neq p, \label{P1_cons:e}\\
		&& \hspace{-5mm} \bm r_k \in\mathcal C^{\rm r}_k, \ \forall k\in\mathcal K, \label{P1_cons:f}
	\end{eqnarray}
\end{subequations}
where $\gamma_k$ in \eqref{P1_cons:b} is the weighting factor controlling the service priority of user $k$, $P_{\max} > 0$ in \eqref{P1_cons:c} represents the maximum instantaneous transmit power of the BS, and \eqref{P1_cons:e} ensures that the distance between each pair of transmit MAs is no smaller than $D > 0$ to avoid the coupling effect between them. We note that $\{\boldsymbol w_n\}$, $\{\boldsymbol t_m\}$, $\{\boldsymbol r_k\}$ are intricately coupled in constraint \eqref{P1_cons:b}. This, along with the non-convex minimum distance constraint \eqref{P1_cons:e}, renders (P1) to be a non-convex optimization problem that presents challenges in seeking an optimal solution. To tackle this issue, we first address the single-group scenario, laying the groundwork for handling the more complex multi-group scenario. Specifically, in the next section, we propose an efficient algorithm to solve (P1) for the single-group case, which is then extended to the multi-group case in Section \ref{Sec:Multi}. 

\section{Single-Group Scenario}\label{Sec:Single}
In this section, we focus on the single-group scenario, i.e., $N = 1$, and omit the group index $n$ for brevity. In this case, due to the absence of inter-group interference, (P1) can be simplified as
\begin{subequations}
	\begin{eqnarray}
		\text{(P2)}: \hspace{-3mm} &\underset{\substack{\eta, \boldsymbol w, \\ \{\boldsymbol t_m\}, \{\boldsymbol r_k\}}}{\max}&  \eta\\
		&\text{s.t.}& \hspace{-5mm}\frac{1}{\gamma_k}\frac{\left|\bm h_k(\bm t, \bm r_k)^H\bm w\right|^2}{\sigma_k^2} \geq \eta, \ \forall k\in\mathcal K, \label{P2_cons:b}\\
		&& \hspace{-5mm} \left\|\bm w\right\|^2 \leq P_{\max}, \label{P2_cons:c}\\
		&& \hspace{-5mm} \eqref{P1_cons:d}-\eqref{P1_cons:f}. 
	\end{eqnarray}
\end{subequations}
While (P2) is much simplified compared to (P1), it remains a non-convex optimization problem because the left-hand-side (LHS) of \eqref{P2_cons:b} is not jointly concave with respect to (w.r.t.) $\left\lbrace \boldsymbol w, \{\boldsymbol t_m\}, \{\boldsymbol r_k\}\right\rbrace$, and the non-convex constraint \eqref{P1_cons:e} still exists. In the following, we explore an efficient AO-based method to address (P2) featuring coupled optimization variables. 

\subsection{Optimizing $\bm w$ for Given $\left\lbrace \{\bm t_m\},\{\bm r_k\}\right\rbrace$}\label{Sec:Single_sub1}
When all the MA positions are given, (P2) reduces to the classical max-min fair beamforming problem in single-group multicast systems, as follows: 
\begin{subequations}\label{P2_sub1}
	\begin{eqnarray}
		&\underset{\eta,\boldsymbol w}{\max}& \eta \\
		&\text{s.t.}& \hspace{-2mm}\bm w^H\mathbf H_k(\bm t, \bm r_k)\bm w \geq \eta\gamma_k\sigma_k^2, \ \forall k\in\mathcal K, \label{P2_sub1_cons:b}\\
		&& \hspace{-2mm} \eqref{P2_cons:c},
	\end{eqnarray}
\end{subequations}
where $\mathbf H_k(\bm t, \bm r_k) \triangleq \bm h_k(\bm t, \bm r_k)\bm h_k(\bm t, \bm r_k)^H$. Although problem \eqref{P2_sub1} is non-convex due to the convexity of $\bm w^H\mathbf H_k(\bm t, \bm r_k)\bm w$ in \eqref{P2_sub1_cons:b}, it can be efficiently solved by using the SCA technique. To elaborate, by replacing $\bm w^H\mathbf H_k(\bm t, \bm r_k)\bm w$ with its global under-estimator based on the first-order Taylor expansion, problem \eqref{P2_sub1} can be approximated as 
\begin{subequations}\label{P2_sub1_sca}
	\begin{eqnarray}
		&\underset{\eta,\boldsymbol w}{\max}& \eta\\
		&\text{s.t.}& \hspace{-2mm} 2\text{Re}\left\lbrace\left( \bm w^\ell\right)^H\mathbf H_k(\bm t, \bm r_k)\bm w\right\rbrace - \left(\bm w^\ell\right)^H\mathbf H_k(\bm t, \bm r_k)\bm w^\ell \nonumber\\
		&& \hspace{-2mm} \geq \eta\gamma_k\sigma_k^2, \ \forall k\in\mathcal K,\\
		&&  \hspace{-2mm} \eqref{P2_cons:c},
	\end{eqnarray}
\end{subequations}
where $\bm w^{\ell}$ denotes the given local point in the $\ell$-th iteration. Problem \eqref{P2_sub1_sca} is a convex quadratically constrained program (QCP) that can be optimally solved utilizing existing solvers like CVX \cite{2004_S.Boyd_cvx}. 

\vspace{-3mm}
\subsection{Optimizing $\bm t_m$ for Given $\left\lbrace \bm w, \{\bm t_p\}_{p\in\mathcal M\setminus\{m\}}, \{\bm r_k\}\right\rbrace$}\label{Sec:Single_sub2}
For any given $\left\lbrace \bm w, \{\bm t_p\}_{p\in\mathcal M\setminus\{m\}}, \{\bm r_k\}\right\rbrace$, the subproblem of (P2) for optimizing $\bm t_m$ is given by
\begin{subequations}\label{P2_sub2}
	\begin{eqnarray}
	    &\underset{\eta, \boldsymbol t_m}{\max}&  \eta\\
		&\text{s.t.}& \hspace{-2mm}\left|\bm h_k(\bm t, \bm r_k)^H\bm w\right|^2 \geq \eta\gamma_k\sigma_k^2, \ \forall k\in\mathcal K, \label{P2_sub2_cons:b}\\
		&& \hspace{-2mm} \bm t_m \in\mathcal C^{\rm t}, \label{P2_sub2_cons:c}\\
		&& \hspace{-2mm} \left\|\bm t_m - \bm t_p\right\| \geq D, \ \forall p\in\mathcal M, p\neq m. \label{P2_sub2_cons:d}
	\end{eqnarray}
\end{subequations}
We notice that the optimization variable $\bm t_m$ is not exposed in the current form of constraint \eqref{P2_sub2_cons:b}. Recall that $\bm h_k(\bm t, \bm r_k)^H = \bm f_k(\bm r_k)^H\mathbf \Sigma_k\bm G_k(\bm t)$. To facilitate the solution design, we define $\bm b_k^H \triangleq \bm f_k(\bm r_k)^H\mathbf \Sigma_k \in\mathbb C^{1\times L_k^{\rm t}}$ and expand the LHS of constraint \eqref{P2_sub2_cons:b} in \eqref{P2_sub2_expand}, as shown at the bottom of the next page.
\begin{figure*}[hb]
	\hrulefill
\begin{align}\label{P2_sub2_expand}
	\left| \bm h_k(\bm t, \bm r_k)^H\bm w\right|^2  = & \left|\bm b_k^H\bm G_k(\bm t)\bm w\right|^2 = \Big|\left[\bm b_k^H\bm g_k(\bm t_1), \cdots, \bm b_k^H\bm g_k(\bm t_M)\right]\bm w\Big|^2 = \left|\sum_{m=1}^M\bm b_k^H\bm g_k(\bm t_m)w_m\right|^2 \nonumber\\
	= & \left(\bm b_k^H\bm g_k(\bm t_m)w_m + \Lambda_{k,m}\right)\left(w_m^*\bm g_k(\bm t_m)^H\bm b_k + \Lambda_{k,m}^*\right) \nonumber\\
	= & \left|w_m\right|^2\bm g_k(\bm t_m)^H\bm B_k\bm g_k(\bm t_m) + 2\text{Re}\left\lbrace w_m\Lambda_{k,m}^*\bm b_k^H\bm g_k(\bm t_m)\right\rbrace + \left| \Lambda_{k,m}\right|^2 \nonumber\\
	= & \sum_{i = 1}^{L_k^{\rm t}-1}\sum_{j=i+1}^{L_k^{\rm t}}2\left|w_m\right|^2\left|\left[ \bm B_k\right]_{i,j} \right|\cos\left(\frac{2\pi}{\lambda}\boldsymbol t_m^T\left(-\boldsymbol a_{k,i}^{\rm t} + \boldsymbol a_{k,j}^{\rm t}\right) + \arg\left(\left[ \bm B_k\right]_{i,j}\right) \right) + \sum_{i=1}^{L_k^{\rm t}}\left|w_m\right|^2\left[ \bm B_k\right]_{i,i}   \nonumber\\
	& + \sum_{i = 1}^{L_k^{\rm t}}2\left|w_m\right|\left|\Lambda_{k,m}\right|\left|\left[ \bm b_k\right]_i \right|\cos\left(\frac{2\pi}{\lambda}\boldsymbol t_m^T\boldsymbol a_{k,i}^{\rm t} + \arg\left( w_m\right)  - \arg\left( \Lambda_{k,m}\right) - \arg \left( \left[\bm b_k\right]_i \right) \right) + \left| \Lambda_{k,m}\right|^2 \nonumber\\
	= & \sum_{i = 1}^{L_k^{\rm t}-1}\sum_{j=i+1}^{L_k^{\rm t}}2\left|w_m\right|^2\left|\left[ \bm B_k\right]_{i,j}\right|\cos\left(\beta_{i,j,k}\left(\bm t_m\right)\right) + \sum_{i=1}^{L_k^{\rm t}}\left|w_m\right|^2\left[ \bm B_k\right]_{i,i} \nonumber\\
	& + \sum_{i = 1}^{L_k^{\rm t}}2\left|w_m\right|\left|\Lambda_{k,m}\right|\left|\left[\bm b_k\right]_i \right|\cos\left(\iota_{k,i}\left(\bm t_m\right)\right)   + \left| \Lambda_{k,m}\right|^2 \nonumber\\
	\triangleq & \; u_k\left(\bm t_m\right). 
\end{align}
\end{figure*}
In \eqref{P2_sub2_expand}, $w_m$ is the $m$-th element of $\bm w$, $\Lambda_{k,m} \triangleq \sum_{p = 1, p\neq m}^M\bm b_k^H\bm g_k(\bm t_p)w_p$, $\bm B_k \triangleq \bm b_k\bm b_k^H$, $\beta_{i,j,k}\left(\bm t_m\right) \triangleq \frac{2\pi}{\lambda}\boldsymbol t_m^T\left(-\boldsymbol a_{k,i}^{\rm t} + \boldsymbol a_{k,j}^{\rm t}\right) + \arg \left(\left[\bm B_k\right]_{i,j}\right)$, and $\iota_{k,i}\left(\bm t_m\right) \triangleq  \frac{2\pi}{\lambda}\boldsymbol t_m^T\boldsymbol a_{k,i}^{\rm t} + \arg\left(w_m\right) - \arg\left(\Lambda_{k,m}\right) - \arg\left( \left[\bm b_k\right]_i\right)$. With \eqref{P2_sub2_expand}, constraint \eqref{P2_sub2_cons:b} can be recast as
\begin{align}\label{P2_sub2_cons:b_eqv} 	u_k\left(\bm t_m\right) \geq \eta\gamma_k\sigma_k^2, \ \forall k\in\mathcal K. 
\end{align} 
However, $u_k\left(\bm t_m\right)$ is neither concave nor convex w.r.t. $\bm t_m$, making constraint \eqref{P2_sub2_cons:b_eqv} still non-convex and also preventing us from constructing a global lower bound for $u_k\left(\bm t_m\right)$ based on its first-order Taylor expansion. To handle \eqref{P2_sub2_cons:b_eqv}, we construct a concave lower-bound surrogate function for $u_k\left(\bm t_m\right)$ by applying the second-order Taylor expansion. Specifically, with a positive real number $\psi_{k,m}$ such that $\psi_{k,m}\mathbf I \succeq \nabla^2 u_k(\bm t_m)$, the following inequality holds: 
\begin{align}\label{ineq:surrogate_func}
	u_k(\bm t_m) & \geq  u_k(\bm t_m^\ell) + \nabla u_k(\bm t_m^\ell)^T(\bm t_m - \bm t_m^\ell) \nonumber\\
	& \hspace{4mm} - \frac{\psi_{k,m}}{2}(\bm t_m - \bm t_m^\ell)^T(\bm t_m - \bm t_m^\ell) \nonumber\\
	& \triangleq u^{\rm lb,\it \ell}_k(\bm t_m),
\end{align}
which is obtained by modifying \cite[Lemma 12]{2017_Sun_MM}. In \eqref{ineq:surrogate_func}, $\bm t_m^\ell$ is the given local point in the $\ell$-th iteration and $\nabla u_k(\bm t_m^\ell)^T = \left[\frac{\partial u_k(\boldsymbol t_m)}{\partial x_m^{\rm t}}\Big|_{\bm t_m = \bm t_m^\ell}, \frac{\partial u_k(\boldsymbol t_m)}{\partial y_m^{\rm t}}\Big|_{\bm t_m = \bm t_m^\ell}\right]$ with the expressions of $\frac{\partial u_k(\boldsymbol t_m)}{\partial x_m^{\rm t}}\Big|_{\bm t_m = \bm t_m^\ell}$ and $\frac{\partial u_k(\boldsymbol t_m)}{\partial y_m^{\rm t}}\Big|_{\bm t_m = \bm t_m^\ell}$ given in \eqref{eq:first-order} at the bottom of this page. 
\begin{figure*}[!hb]
	\hrulefill
\begin{subequations}\label{eq:first-order}
	\begin{align}
	 \frac{\partial u_k(\boldsymbol t_m)}{\partial x_m^{\rm t}}\Big|_{\bm t_m = \bm t_m^\ell} = & -\frac{4\pi}{\lambda}\sum_{i = 1}^{L_k^{\rm t}-1}\sum_{j = i+1}^{L_k^{\rm t}}\left|w_m \right|^2\left|\left[ \bm B_k\right]_{i,j}\right|\left(- \cos\theta_{k,i}^{\rm t}\sin\phi_{k,i}^{\rm t} + \cos\theta_{k,j}^{\rm t}\sin\phi_{k,j}^{\rm t}\right)\sin\left(\beta_{i,j,k}(\bm t_m^\ell)\right) \nonumber\\
	 & -\frac{4\pi}{\lambda}\sum_{i = 1}^{L_k^{\rm t}}\left|w_m\right|\left|\Lambda_{k,m}\right|\left|\left[ \bm b_k\right]_i \right|\cos\theta_{k,i}^{\rm t}\sin\phi_{k,i}^{\rm t}\sin\left(\iota_{k,i}(\bm t_m^\ell)\right), \\
	 \frac{\partial u_k(\boldsymbol t_m)}{\partial y_m^{\rm t}}\Big|_{\bm t_m = \bm t_m^\ell} = & -\frac{4\pi}{\lambda}\sum_{i = 1}^{L_k^{\rm t}-1}\sum_{j = i+1}^{L_k^{\rm t}}\left|w_m \right|^2\left|\left[\bm B_k\right]_{i,j} \right|\left(-\sin\theta_{k,i}^{\rm t} + \sin\theta_{k,j}^{\rm t}\right)\sin\left(\beta_{i,j,k}(\bm t_m^\ell)\right) \nonumber\\
	 & -\frac{4\pi}{\lambda}\sum_{i = 1}^{L_k^{\rm t}}\left|w_m\right|\left|\Lambda_{k,m}\right|\left|\left[\bm b_k\right]_i \right|\sin\theta_{k,i}^{\rm t}\sin\left(\iota_{k,i}(\bm t_m^\ell)\right).
	\end{align}
\end{subequations}
\end{figure*}
How to construct a $\psi_{k,m}$ is detailed in the appendix. 

With \eqref{ineq:surrogate_func}, a convex subset of constraint \eqref{P2_sub2_cons:b_eqv} is given by
\begin{align}\label{P2_sub2_cons:b_eqv_sca} 
	u^{\rm lb,\it \ell}_k(\bm t_m) \geq \eta\gamma_k\sigma_k^2, \ \forall k\in\mathcal K. 
\end{align}
The sole remaining hurdle to solving problem \eqref{P2_sub2} lies in the non-convex constraint \eqref{P2_sub2_cons:d}. Note that constraint \eqref{P2_sub2_cons:d} is equivalent to $\left\|\bm t_m - \bm t_p\right\|^2 \geq D^2, \forall p\in\mathcal M, p\neq m$. Since the term $\left\|\bm t_m - \bm t_p\right\|^2$ is convex w.r.t. $\bm t_m$, it is lower bounded by its first-order Taylor expansion, i.e., 
\begin{align}\label{ineq:t_Taylor}
	\left\|\bm t_m - \bm t_p\right\|^2 & \geq \left\|\bm t_m^\ell - \bm t_p\right\|^2 + 2\left(\bm t_m^\ell - \bm t_p\right)^T\left(\bm t_m - \bm t_m^\ell\right) \nonumber\\
	& \triangleq \mathcal T^{\rm lb, \it\ell}(\bm t_m).
\end{align}
By replacing $\left\|\bm t_m - \bm t_p\right\|^2$ with $\mathcal T^{\rm lb, \it\ell}(\bm t_m)$ and constraint \eqref{P2_sub2_cons:b} with \eqref{P2_sub2_cons:b_eqv_sca}, we can acquire a lower bound of the optimal value of problem \eqref{P2_sub2} by solving
\begin{subequations}\label{P2_sub2_sca}
	\begin{eqnarray}
		&\underset{\eta, \boldsymbol t_m}{\max}&  \eta\\
		&\text{s.t.}& \hspace{-2mm}  \eqref{P2_sub2_cons:c}, \eqref{P2_sub2_cons:b_eqv_sca},\\
		&& \hspace{-2mm} \mathcal T^{\rm lb, \it\ell}(\bm t_m) \geq D^2, \ \forall p\in\mathcal M, p\neq m.  
	\end{eqnarray}
\end{subequations}
As a convex QCP, this problem's optimal solution can be obtained readily by standard solvers (e.g., CVX \cite{2004_S.Boyd_cvx}). 

\begin{rem}\label{rem1}
	\rm It is worth mentioning that after expanding $\left| \bm h_k(\bm t, \bm r_k)^H\bm w\right|^2$ as $\left| \bm h_k(\bm t, \bm r_k)^H\bm w\right|^2 = \left|w_m\right|^2\bm g_k(\bm t_m)^H\bm B_k\bm g_k(\bm t_m) + 2\text{Re}\left\lbrace w_m\Lambda_{k,m}^*\bm b_k^H\bm g_k(\bm t_m)\right\rbrace + \left| \Lambda_{k,m}\right|^2$, if one applies the two-step approximation method proposed in \cite{2023_Wenyan_MIMO} to handle this expansion, one needs to first apply the first-order Taylor expansion to the term  $\left|w_m\right|^2\bm g_k(\bm t_m)^H\bm B_k\bm g_k(\bm t_m)$ to obtain its lower bound, followed by utilizing the second-order Taylor expansion to construct surrogate functions for the resulting lower bound and the term 2$\text{Re}\left\lbrace w_m\Lambda_{k,m}^*\bm b_k^H\bm g_k(\bm t_m)\right\rbrace$. By contrast, our proposed method requires only one approximation to obtain a surrogate lower bound of $\left|\bm h_k(\bm t, \bm r_k)^H\bm w\right|^2$, i.e., $u^{\rm lb,\it \ell}_k(\bm t_m)$. Compared with the method in \cite{2023_Wenyan_MIMO}, our proposed method not only simplifies the problem-solving process but is also expected to be more effective in terms of the converged solution, which is validated later by the simulation results in Fig. \ref{fig:single_vs_iter} in Section \ref{Sec:Simulation}. 
\end{rem}  

\subsection{Optimizing $\{\bm r_k\}$ for Given $\left\lbrace \bm w, \{\bm t_m\}\right\rbrace$}\label{Sec:single_sub3}
Observe that unlike $\{\bm t_m\}$, each $\bm r_k$ appears exclusively in its own achievable SNR without mutual coupling. Thus, all users can achieve their individual maximum SNR simultaneously via independently optimizing each $\bm r_k$, thereby maximizing the minimum SNR among them. Specifically, $\{\bm r_k\}$ can be independently optimized by solving $K$ subproblems in parallel, each with only one position variable, as described below:
\begin{subequations}\label{P2_sub3}
	\begin{eqnarray}
		&\underset{\boldsymbol r_k}{\max}&  \frac{1}{\gamma_k}\frac{\left|\bm h_k(\bm t, \bm r_k)^H\bm w\right|^2}{\sigma_k^2}\\ \label{P2_sub3_obj}
		&\text{s.t.}& \hspace{-2mm} \bm r_k \in\mathcal C^{\rm r}_k, \ k\in\mathcal K. \label{P2_sub3_cons:b}
	\end{eqnarray}
\end{subequations}
In order to expose $\bm r_k$ in \eqref{P2_sub3_obj}, we expand $\left|\bm h_k(\bm t, \bm r_k)^H\bm w\right|^2$ as follows:
\begin{align}\label{eq:r_expand_single}
	& \left|\bm h_k(\bm t, \bm r_k)^H\bm w\right|^2  \nonumber\\
	& = \bm f_k(\bm r_k)^H\mathbf \Sigma_k\bm G_k(\bm t)\bm w\bm w^H\bm G_k(\bm t)^H\mathbf \Sigma_k^H\bm f_k(\bm r_k) \nonumber\\
	& = \bm f_k(\bm r_k)^H\bm C_k\bm f_k(\bm r_k) \nonumber\\
	& = \sum_{i = 1}^{L_k^{\rm r}-1}\sum_{j=i+1}^{L_k^{\rm r}}2\left|\left[\bm C_k\right]_{i,j} \right|\cos\left(\tau_{i,j}\left( \bm r_k\right) \right) + \sum_{i = 1}^{L_k^{\rm r}}\left[ \bm C_k\right]_{i,i}\nonumber\\
	& \triangleq v\left(\bm r_k\right), 
\end{align}
where $\bm C_k \triangleq \mathbf \Sigma_k\bm G_k(\bm t)\bm w\bm w^H\bm G_k(\bm t)^H\mathbf \Sigma_k^H \in \mathbb C^{L_k^{\rm r} \times L_k^{\rm r}}$ and $\tau_{i,j}\left( \bm r_k\right) \triangleq \frac{2\pi}{\lambda}\boldsymbol r_k^T\left(-\boldsymbol a_{k,i}^{\rm r} + \boldsymbol a_{k,j}^{\rm r}\right) + \arg\left(\left[\bm C_k\right]_{i,j}\right)$. Note that $\bm C_k$ is neither concave nor convex w.r.t. $\bm r_k$. Similar to \eqref{ineq:surrogate_func}, we can construct a second-order Taylor expansion-based concave lower bound for $v\left(\bm r_k\right)$, denoted by $v^{\rm lb,\it \ell}(\bm r_k) \triangleq v(\bm r_k^\ell) + \nabla v(\bm r_k^\ell)^T(\bm r_k - \bm r_k^\ell) - \frac{\hat \psi_k}{2}(\bm r_k - \bm r_k^\ell)^T(\bm r_k - \bm r_k^\ell)$. We then replace $\left|\bm h_k(\bm t, \bm r_k)^H\bm w\right|^2$ in \eqref{P2_sub3_obj} with $v^{\rm lb,\it \ell}(\bm r_k)$, which yields the following QCP:
\begin{subequations}\label{P2_sub3_sca}
	\begin{eqnarray}
		&\underset{\boldsymbol r_k}{\max}&  \frac{1}{\gamma_k}\frac{v^{\rm lb,\it \ell}(\bm r_k)}{\sigma_k^2}\\
		&\text{s.t.}& \eqref{P2_sub3_cons:b}. 
	\end{eqnarray}
\end{subequations}
If ignoring constraint \eqref{P2_sub3_cons:b}, the solution for maximizing $v^{\rm lb,\it \ell}(\bm r_k)$ can be obtained in a closed-form expression as $\bm r_k^* = \frac{\nabla v(\bm r_k^\ell)}{\hat \psi_k} + \bm r_k^\ell$. If $\bm r_k^*$ satisfies \eqref{P2_sub3_cons:b}, it is the optimal solution of problem \eqref{P2_sub3_sca}. Otherwise, we utilize off-the-shelf solvers, such as CVX \cite{2004_S.Boyd_cvx}, to solve problem \eqref{P2_sub3_sca} optimally. 

Note that if computational resources do not allow solving $K$ subproblems of the form \eqref{P2_sub3_sca} in parallel, we can either solve them one by one or collectively updating $\{\bm r_k\}$ by solving the following single problem:
\begin{subequations}\label{P2_sub3_sca_all}
	\begin{eqnarray}
		&\underset{\eta,\{\boldsymbol r_k\}}{\max}& \eta \\
		&\text{s.t.}& \frac{1}{\gamma_k}\frac{v^{\rm lb,\it \ell}(\bm r_k)}{\sigma_k^2} \geq \eta, \ \forall k\in\mathcal K, \\
		&& \bm r_k \in\mathcal C^{\rm r}_k, \ \forall k\in\mathcal K. 
	\end{eqnarray}
\end{subequations}
These three ways of updating $\{\bm r_k\}$ yield the same max-min SNR.

\subsection{Overall Algorithm: Convergence and Complexity Analysis}
\begin{algorithm}[!t]  
	\caption{Proposed AO-based algorithm for problem (P2)}  \label{Alg1}  
	\begin{algorithmic}[1]
		\STATE Initialize $\left\lbrace \boldsymbol w^0, \{\boldsymbol t_m^0\}, \{\boldsymbol r_k^0\}\right\rbrace$ and set $\ell = 0$.  
		\REPEAT 
		\STATE Obtain $\bm w^{\ell + 1}$ by solving problem \eqref{P2_sub1_sca} for given $\left\lbrace \{\bm t_m^\ell\},\{\bm r_k^\ell\}\right\rbrace$. 
		\FOR{$m = 1$ to $M$}
		\STATE Compute $\{\bm b_k, \bm B_k, \Lambda_{k,m}\}_{k=1}^K$. 
		\STATE Compute $\{\nabla u_k(\bm t_m^\ell), \nabla u_k^2(\bm t_m), \psi_{k,m}\}_{k=1}^K$ via \eqref{eq:first-order}, \eqref{eq:second-order}, and \eqref{ineq:psi}. 
		\STATE Obtain $\bm t_m^{\ell + 1}$ by solving problem \eqref{P2_sub2_sca}, given $\left\lbrace \bm w^{\ell + 1}, \bm t_1^{\ell+1},\cdots,\bm t_{m-1}^{\ell+1}, \bm t_m^\ell,\cdots, \bm t_M^\ell, \{\bm r_k^\ell\}\right\rbrace$.  
		\ENDFOR
		\STATE Compute $\{\bm C_k\}_{k=1}^K$.
		\STATE Compute $\{\nabla v(\bm r_k^\ell), \nabla v^2(\bm r_k), \hat\psi_k\}_{k=1}^K$. 
		\STATE Obtain $\{\bm r_k^{\ell + 1}\}$ by solving $K$ problems of the form \eqref{P2_sub3_sca} one by one or in parallel (if computational resources allow) or by solving only problem \eqref{P2_sub3_sca_all} for given $\left\lbrace \bm w^{\ell + 1}, \{\bm t_m^{\ell + 1}\}, \{\bm r_k^\ell\}\right\rbrace$.
		\STATE $\ell \leftarrow \ell + 1$.
		\UNTIL The fractional increase of the objective value of problem (P2) between two consecutive iterations drops below a threshold $\epsilon > 0$. 
	\end{algorithmic} 
\end{algorithm}

Based on the above results, we outline our proposed algorithm for (P2) in Algorithm \ref{Alg1}. It can be proved as in \cite{2018_Qingqing_multiUAV} that repeating steps 3-12 of Algorithm \ref{Alg1} yields a non-decreasing sequence of objective values of (P2). Furthermore, the optimal value of (P2) is upper-bounded. As a result, Algorithm \ref{Alg1} is assured to converge at a suboptimal solution. Next, we analyze the complexity of this algorithm. In step 3, the computational cost of updating $\bm w$ via solving problem \eqref{P2_sub1_sca} is $\mathcal O\left(\ln\frac{1}{\varepsilon} K^{1.5}M^{4.5}\right)$, where $\varepsilon$ is the prescribed accuracy \cite{2014_K.wang_complexity}. The complexities of steps 5-7 are $\mathcal O\left(K\left(L_k^{\rm t}L_k^{\rm r} + \left(L_k^{\rm t}\right)^2\right)\right)$, $\mathcal O\left(K\left(L_k^{\rm t}\right)^2\right)$, and $\mathcal O\left(\ln\frac{1}{\varepsilon}\left(K+M \right)^{1.5}\right)$, respectively. Thus, the complexity of updating $\{\bm t_m\}$ by executing steps 4-8 is $\mathcal O\left(M\left(K\left(L_k^{\rm t}L_k^{\rm r} + \left(L_k^{\rm t}\right)^2\right) + \ln\frac{1}{\varepsilon}\left(K+M \right)^{1.5}\right) \right)$. Similarly, the complexity of updating $\{\bm r_k\}$ by executing steps 9-11 is $\mathcal O\left(K\left(\left(L_k^{\rm r}\right)^2 + L_k^{\rm r}L_k^{\rm t} + L_k^{\rm t}M\right) \right)$. Therefore,  the computational complexity of each iteration of Algorithm \ref{Alg1} is $\mathcal O\Big(\ln\frac{1}{\varepsilon} K^{1.5}M^{4.5} + MK\left(L_k^{\rm t}L_k^{\rm r} + \left(L_k^{\rm t}\right)^2\right) + M\ln\frac{1}{\varepsilon}\left(K+M \right)^{1.5} + K\left(L_k^{\rm r}\right)^2\Big)$. 

\begin{rem}\label{rem2}
	\rm Notably, compared with the two-step approximation method proposed in \cite{2023_Wenyan_MIMO}, although our one-step approximation algorithm simplifies the problem-solving process, it does not reduce the order of computational complexity. Nevertheless, simulation results show that our algorithm converges more quickly and offers a 3.4\% improvement in max-min SNR when the convergence threshold is set to $10^{-4}$. Moreover, in \cite{2023_Wenyan_MIMO}, the positions of receive MAs can only be updated one by one. In contrast, our algorithm allows for sequential, independent parallel, or collective updates of receive MA positions, providing greater flexibility and adaptability during execution. If computational resources allow parallel computation, the computational time will be significantly reduced compared to the one-by-one update, especially for large $K$.
	
	\rm Additionally, it should be noted that even with the exhaustive search method, the optimal solution of (P2) cannot be obtained to evaluate the performance gap with our suboptimal solution. The reasons are as follows. First, the number of candidate positions for each MA is infinite due to the continuity of the moving regions, making it impossible to exhaustively search all potential positions. If one intends to use the exhaustive search method, it is necessary to quantify the movement regions and model the motion of the MAs as discrete steps. Assuming each MA has $Q$ candidate positions, the complexity of using the exhaustive search method to find the optimal positions for $M$ transmit MAs and $K$ receive MAs is $\mathcal O\left(Q^{M+N}\right)$. This exponential complexity is significantly higher than the polynomial complexity of our proposed algorithm, and the obtained solution is generally not optimal for the original problem. Second, given the MA positions, the remaining problem with respect to the transmit beamforming is still NP-hard and cannot be directly solved via existing optimization techniques.
\end{rem}  

\section{Multi-Group Scenario}\label{Sec:Multi}
In this section, we consider the general multi-group multicast scenario and extend the AO-based algorithm proposed in the previous section to address the corresponding problem in this challenging setup. 

\subsection{Optimizing $\{\bm w_n\}$ for Given $\left\lbrace \{\bm t_m\}, \{\bm r_k\}\right\rbrace$} 
Given any fixed $\left\lbrace \{\bm t_m\}, \{\bm r_k\}\right\rbrace$, and recalling that we defined $\mathbf H_k(\bm t, \bm r_k) \triangleq \bm h_k(\bm t, \bm r_k)\bm h_k(\bm t, \bm r_k)^H$ in Section \ref{Sec:Single_sub1}, the subproblem of (P1) regarding to $\{\bm w_n\}$ can be expressed as
\begin{subequations}\label{P1_sub1}
	\begin{eqnarray}
		&\hspace{-3mm}\underset{\eta, \{\boldsymbol w_n\}}{\max}& \eta\\
		&\hspace{-6.5mm}\text{s.t.}& \hspace{-6.5mm} \frac{1}{\gamma_k}\frac{\bm w_n^H\mathbf H_k(\bm t, \bm r_k)\bm w_n}{\eta} \geq \sum_{q=1,q\neq n}^N\bm w_q^H\mathbf H_k(\bm t, \bm r_k)\bm w_q + \sigma_k^2, \nonumber\\
		&& \hspace{-6.5mm} \forall k\in\mathcal G_n,  n\in\mathcal N, \label{P1_sub1_cons:b}\\
		&& \hspace{-6.5mm} \eqref{P1_cons:c}.
	\end{eqnarray}
\end{subequations}
Note that the convexity of $\frac{\bm w_n^H\mathbf H_k(\bm t, \bm r_k)\bm w_n}{\eta}$ results in the non-convexity of constraint \eqref{P1_sub1_cons:b}. This prompts us to approximate $\frac{\bm w_n^H\mathbf H_k(\bm t, \bm r_k)\bm w_n}{\eta}$ by its first-order Taylor expansion-based affine under-estimator, yielding 
\begin{subequations}\label{P1_sub1_sca}
	\begin{eqnarray}
		&\underset{\eta, \{\boldsymbol w_n\}}{\max}&  \eta\\
		&\text{s.t.}& \hspace{-3mm} \frac{1}{\gamma_k}\mathcal F^{\rm lb, \it r}\left(\bm w_n\right) \geq \sum_{q=1,q\neq n}^N\bm w_q^H\mathbf H_k(\bm t, \bm r_k)\bm w_q + \sigma_k^2, \nonumber\\
		&& \hspace{-3mm} \forall k\in\mathcal G_n, n\in\mathcal N, \\
		&& \hspace{-3mm} \eqref{P1_cons:c},
	\end{eqnarray}
\end{subequations}
where
\begin{align}
\mathcal F^{\rm lb, \it r}\left(\bm w_n\right) \triangleq & \frac{2\text{Re}\left\lbrace \left(\bm w_n^r\right)^H\mathbf H_k(\bm t, \bm r_k)\bm w_n\right\rbrace}{\eta^r} \nonumber\\
& -\frac{\left(\bm w_n^r\right)^H\mathbf H_k(\bm t, \bm r_k)\bm w_n^r}{\left( \eta^r\right)^2}\eta
\end{align}
with $\bm w_n^r$ and $\eta^r$ being the given local points in the $r$-th iteration. By solving the convex QCP in \eqref{P1_sub1_sca} via standard solvers such as CVX \cite{2004_S.Boyd_cvx}, we obtain a performance lower bound of problem \eqref{P1_sub1}. 

\subsection{Optimizing $\bm t_m$ for Given $\left\lbrace \{\bm w_n\}, \{\bm t_p\}_{p\in\mathcal M\setminus\{m\}}, \{\bm r_k\}\right\rbrace$}
With fixed $\left\lbrace \{\bm w_n\}, \{\bm t_p\}_{p\in\mathcal M\setminus\{m\}}, \{\bm r_k\}\right\rbrace$, (P1) reduces to
\begin{subequations}\label{P1_sub2}
	\begin{eqnarray}
		&\underset{\eta, \boldsymbol t_m}{\max}&  \eta\\
		&\text{s.t.}& \hspace{-2mm} \frac{1}{\gamma_k}\frac{\left|\bm h_k(\bm t, \bm r_k)^H\bm w_n\right|^2}{\sum_{q=1,q\neq n}^N\left|\bm h_k(\bm t, \bm r_k)^H\bm w_q\right|^2 + \sigma_k^2} \geq \eta, \nonumber\\
		&& \hspace{-2mm} \forall k\in\mathcal G_n, n\in\mathcal N, \label{P1_sub2_cons:b}\\
		&& \hspace{-2mm} \bm t_m \in\mathcal C^{\rm t}, \label{P1_sub2_cons:c}\\
		&& \hspace{-2mm} \left\|\bm t_m - \bm t_p\right\| \geq D, \ \forall p\in\mathcal M, p\neq m. \label{P1_sub2_cons:d}  
	\end{eqnarray}
\end{subequations}
Similar to \eqref{P2_sub2_expand}, $\left|\bm h_k(\bm t, \bm r_k)^H\bm w_n\right|^2$ in constraint \eqref{P1_sub2_cons:b} can be expanded as
\begin{align}\label{P1_sub2_expand}
	& \hspace{-5mm}\left| \bm h_k(\bm t, \bm r_k)^H\bm w_n\right|^2\nonumber\\
	& = \sum_{i = 1}^{L_k^{\rm t}-1}\sum_{j=i+1}^{L_k^{\rm t}}2\left|w_{n,m}\right|^2\left|\left[\bm B_k\right]_{i,j}\right|\cos\left(\beta_{i,j,k}\left(\bm t_m\right)\right) \nonumber\\ 
	& \hspace{4mm} + \sum_{i=1}^{L_k^{\rm t}}\left|w_{n,m}\right|^2\left[ \bm B_k\right]_{i,i}  + \left| \Lambda_{k,n,m}\right|^2 \nonumber\\
	& \hspace{4mm}  + \sum_{i = 1}^{L_k^{\rm t}}2\left|w_{n,m}\right|\left|\Lambda_{k,n,m}\right|\left|\left[\bm b_k\right]_i \right|\cos\left(\iota_{k,i,n}\left(\bm t_m\right)\right) \nonumber\\
	& \triangleq u_{k,n}\left(\bm t_m\right),
\end{align}
where $\bm B_k$ and $\beta_{i,j,k}\left(\bm t_m\right)$ are defined in Section \ref{Sec:Single_sub2}, $w_{n,m}$ is the $m$-th element of $\bm w_n$, $\Lambda_{k,n,m} \triangleq \sum_{p = 1, p\neq m}^M\bm b_k^H\bm g_k(\bm t_p)w_{n,p}$, and $\iota_{k,i,n}\left(\bm t_m\right) \triangleq  \frac{2\pi}{\lambda}\boldsymbol t_m^T\boldsymbol a_{k,i}^{\rm t} + \arg\left(w_{n,m}\right) - \arg\left(\Lambda_{k,n,m}\right) - \arg\left( \left[\bm b_k\right]_i \right)$. With \eqref{P1_sub2_expand}, constraint \eqref{P1_sub2_cons:b} can be recast as 
\begin{align}\label{P1_sub2_cons:b_eqv1}
	\frac{1}{\gamma_k}\frac{u_{k,n}\left(\bm t_m\right)}{\sum_{q=1,q\neq n}^Nu_{k,q}\left(\bm t_m\right) + \sigma_k^2} \geq \eta, \ \forall k\in\mathcal G_n, n\in\mathcal N,
\end{align}
where the expression of $u_{k,q}\left(\bm t_m\right)$ can be obtained by replacing the index ``$n$'' in \eqref{P1_sub2_expand} with ``$q$''. Note that $u_{k,n}\left(\bm t_m\right)$ and $u_{k,q}\left(\bm t_m\right)$ are neither concave nor convex, and the same applies to $\frac{u_{k,n}\left(\bm t_m\right)}{\eta}$. If we rewrite constraint \eqref{P1_sub2_cons:b_eqv1} as $\frac{1}{\gamma_k}\frac{u_{k,n}\left(\bm t_m\right)}{\eta} \geq \sum_{q=1,q\neq n}^Nu_{k,q}\left(\bm t_m\right) + \sigma_k^2, \forall k\in\mathcal G_n, n\in\mathcal N$, which is in the same form as constraint \eqref{P1_sub1_cons:b} in the previous subsection, the resulting constraint is still intractable. To address this issue, we introduce slack variables $\{z_k\}$ and convert \eqref{P1_sub2_cons:b_eqv1} to
\begin{subequations}\label{P1_cons:b_eqv2}
	\begin{align}
		& \frac{1}{\gamma_k}u_{k,n}\left(\bm t_m\right) \geq \eta z_k,  \ \forall k\in\mathcal G_n, n\in\mathcal N, \label{P1_cons:b_eqv2_1}\\
		& \sum_{q=1,q\neq n}^Nu_{k,q}\left(\bm t_m\right) +  \sigma_k^2 \leq z_k, \ \forall k\in\mathcal G_n, n\in\mathcal N. \label{P1_cons:b_eqv2_2}
	\end{align}
\end{subequations}
The equivalence between \eqref{P1_sub2_cons:b_eqv1} and \eqref{P1_cons:b_eqv2} can be verified by contradiction. However, the constraints in \eqref{P1_cons:b_eqv2} are still non-convex. We first deal with \eqref{P1_cons:b_eqv2_1}. Similar to \eqref{ineq:surrogate_func}, with the given local point $\bm t_m^r$ in the $r$-th iteration, we construct a concave lower-bound surrogate function, denoted by $u_{k,n}^{\rm lb,\it r}(\bm t_m)$, for the LHS of \eqref{P1_cons:b_eqv2_1}. For the right-hand-side of \eqref{P1_cons:b_eqv2_1}, it is not jointly convex w.r.t. $\eta$ and $z_k$ but satisfies \cite[(101)]{2017_Sun_MM}:
\begin{align}
	\eta z_k \leq \frac{1}{2}\left(\frac{z_k^r}{\eta^r}\eta^2 + \frac{\eta^r}{z_k^r}z_k^2 \right) \triangleq \chi^{\rm ub, \it r}\left(\eta, z_k\right),
\end{align}
where $\eta^r$ and $z_k^r$ are the given local points in the $r$-th iteration. Then, a convex approximation of constraint \eqref{P1_cons:b_eqv2_1} can be established immediately, given by
\begin{align}
	\frac{1}{\gamma_k}u_{k,n}^{\rm lb,\it r}(\bm t_m) \geq \chi^{\rm ub, \it r}\left(\eta, z_k\right), \ \forall k\in\mathcal G_n, n\in\mathcal N. \label{P1_cons:b_eqv2_1_sca}
\end{align}
To proceed, we tackle the non-convex constraint \eqref{P1_cons:b_eqv2_2}. As we need to construct an upper bound for $u_{k,q}\left(\bm t_m\right)$, the inequality in \eqref{ineq:surrogate_func} is not applicable. Nevertheless, according to \cite[Lemma 12]{2017_Sun_MM}, we have
\begin{align}\label{ineq:surrogate_func_ub}
	u_{k,q}(\bm t_m) & \leq  u_{k,q}(\bm t_m^r) + \nabla u_{k,q}(\bm t_m^r)^T(\bm t_m - \bm t_m^r) \nonumber\\
	& \hspace{4mm} + \frac{\psi_{k,q,m}}{2}(\bm t_m - \bm t_m^r)^T(\bm t_m - \bm t_m^r) \nonumber\\
	& \triangleq u_{k,q}^{\rm ub,\it r}(\bm t_m),
\end{align} 
where $\psi_{k,q,m}$ is a positive real number satisfying $\psi_{k,q,m}\mathbf I \succeq \nabla^2 u_{k,q}(\bm t_m)$ and can be determined by following similar steps as in \eqref{ineq:psi}. Subsequently, constraint \eqref{P1_cons:b_eqv2_2} can be approximated as \looseness=-1
\begin{align}
	\sum_{q=1,q\neq n}^Nu_{k,q}^{\rm ub,\it r}(\bm t_m) +  \sigma_k^2 \leq z_k, \ \forall k\in\mathcal G_n, n\in\mathcal N, \label{P1_cons:b_eqv2_2_sca}
\end{align}
which is a convex constraint. 

Now, the only barrier left in solving problem \eqref{P1_sub2} is the non-convex constraint \eqref{P1_sub2_cons:d}, which is exactly the same as constraint \eqref{P2_sub2_cons:d}. Recall that in Section \ref{Sec:Single_sub2}, we approximated constraint \eqref{P2_sub2_cons:d} as a convex one, which can be expressed as
\begin{align}\label{P1_sub2_cons:d_sca}
	\mathcal T^{\rm lb, \it r}(\bm t_m) \geq D^2, \ \forall p\in\mathcal M, p\neq m, 
\end{align} 
where the expression of $\mathcal T^{\rm lb, \it r}(\bm t_m)$ can be obtained  by replacing the index ``$\ell$'' in \eqref{ineq:t_Taylor} with ``$r$''. Consequently, by replacing constraint \eqref{P1_sub2_cons:b} with \eqref{P1_cons:b_eqv2_1_sca} and \eqref{P1_cons:b_eqv2_2_sca}, and replacing constraint \eqref{P1_sub2_cons:d} with \eqref{P1_sub2_cons:d_sca}, we arrive at the following problem: 
\begin{subequations}\label{P1_sub2_sca}
	\begin{eqnarray}
		&\underset{\eta, \boldsymbol t_m, \{z_k\}}{\max}&  \eta\\
		&\text{s.t.}& \hspace{-4mm} \eqref{P1_sub2_cons:c}, \eqref{P1_cons:b_eqv2_1_sca}, \eqref{P1_cons:b_eqv2_2_sca}, \eqref{P1_sub2_cons:d_sca}.
	\end{eqnarray}
\end{subequations}
Note that problem \eqref{P1_sub2_sca} is a convex QCP that can be optimally solved using CVX \cite{2004_S.Boyd_cvx}, and its optimal value serves as a lower bound of that of problem \eqref{P1_sub2}.

\subsection{Optimizing $\{\bm r_k\}$ for Given $\left\lbrace \bm w, \{\bm t_m\}\right\rbrace$}
Similar to Section \ref{Sec:single_sub3}, $\{\bm r_k\}$ can be optimized independently by solving $K$ subproblems in parallel, with the $k$-th subproblem given by
\begin{subequations}\label{P1_sub3}
	\begin{eqnarray}
		&\underset{\eta_k, \{\boldsymbol r_k\}}{\max}&  \eta_k\\
		&\text{s.t.}& \hspace{-2mm} \frac{1}{\gamma_k}\frac{\left|\bm h_k(\bm t, \bm r_k)^H\bm w_n\right|^2}{\sum_{q=1,q\neq n}^N\left|\bm h_k(\bm t, \bm r_k)^H\bm w_q\right|^2 + \sigma_k^2} \geq \eta_k, \nonumber\\
		&& \hspace{-2mm} k\in\mathcal G_n, n\in\mathcal N, \label{P1_sub3_cons:b}\\
		&& \hspace{-2mm} \bm r_k \in\mathcal C^{\rm r}_k, \ k\in\mathcal K, \label{P1_sub3_cons:c}  
	\end{eqnarray}
\end{subequations}
As in \eqref{eq:r_expand_single}, we can express the term $\left|\bm h_k(\bm t, \bm r_k)^H\bm w_n\right|^2$ as
\begin{align}\label{eq:r_expand_multi}
	& \left|\bm h_k(\bm t, \bm r_k)^H\bm w_n\right|^2 \nonumber\\
	& = \sum_{i = 1}^{L_k^{\rm r}-1}\sum_{j=i+1}^{L_k^{\rm r}}2\left|\left[\bm C_{k,n}\right]_{i,j} \right|\cos\left(\tau_{i,j,n}\left( \bm r_k\right) \right) + \sum_{i = 1}^{L_k^{\rm r}}\left[ \bm C_{k,n}\right]_{i,i}\nonumber\\
	& \triangleq v_n\left(\bm r_k\right), 
\end{align}
where $\bm C_{k,n} \triangleq \mathbf \Sigma_k\bm G_k(\bm t)\bm w_n\bm w_n^H\bm G_k(\bm t)^H\mathbf \Sigma_k^H \in \mathbb C^{L_k^{\rm r} \times L_k^{\rm r}}$ and $\tau_{i,j,n}\left(\bm r_k\right) \triangleq \frac{2\pi}{\lambda}\boldsymbol r_k^T\left(-\boldsymbol a_{k,i}^{\rm r} + \boldsymbol a_{k,j}^{\rm r}\right) + \arg\left(\left[ \bm C_{k,n}\right] _{i,j}\right)$. Also, replacing the index ``$n$'' with ``$q$'' in \eqref{eq:r_expand_multi}, we can obtain the expansion of $\left|\bm h_k(\bm t, \bm r_k)^H\bm w_q\right|^2$, i.e., $v_q\left(\bm r_k\right)$. Then, constraint \eqref{P1_sub3_cons:b} can be transformed into
\begin{align}\label{P1_sub3_cons:b_eqv1}
	\frac{1}{\gamma_k}\frac{v_n\left(\bm r_k\right)}{\sum_{q=1,q\neq n}^Nv_q\left(\bm r_k\right) + \sigma_k^2} \geq \eta_k, \ k\in\mathcal G_n, n\in\mathcal N. 
\end{align}
Since constraint \eqref{P1_sub3_cons:b_eqv1} shares a similar structure with constraint \eqref{P1_sub2_cons:b_eqv1}, it can be addressed in the same manner as for \eqref{P1_sub2_cons:b_eqv1}, which is thus omitted for brevity. After replacing constraint \eqref{P1_sub3_cons:b} with the convex approximation of constraint \eqref{P1_sub3_cons:b_eqv1}, a lower bound of the optimal value of the subproblem \eqref{P1_sub3} can be obtained by solving the resulting convex QCP.

\subsection{Overall Algorithm: Convergence and Complexity Analysis}
The detailed steps of the proposed algorithm for (P1) are not provided here, as they closely resemble those outlined in Algorithm \ref{Alg1} for (P2). Moreover, the convergence of this algorithm is assured for identical reasons to Algorithm \ref{Alg1}. Besides, in each iteration of this algorithm, the complexity of updating $\{\bm w_n\}$, $\bm t_m$, and $\{\bm r_k\}$ is given by $\mathcal O\left(\ln\frac{1}{\varepsilon} K^{1.5}M^3N^3\right)$, $\mathcal O\left(M\left(KN\left(L_k^{\rm t}L_k^{\rm r} + \left(L_k^{\rm t}\right)^2\right) + \ln\frac{1}{\varepsilon}K^2\left(K+M \right)^{1.5}\right) \right)$, and $\mathcal O\left(KN\left(\left(L_k^{\rm r}\right)^2 + L_k^{\rm r}L_k^{\rm t} + L_k^{\rm t}M\right) \right)$, respectively, with $\varepsilon$ representing the solution accuracy \cite{2014_K.wang_complexity}. Thus, the computational complexity of each iteration of this algorithm is about $\mathcal O\Big(\ln\frac{1}{\varepsilon} K^{1.5}M^3N^3 + MKN\left(L_k^{\rm t}L_k^{\rm r} + \left(L_k^{\rm t}\right)^2\right) + M\ln\frac{1}{\varepsilon}K^2\left(K+M\right)^{1.5} + KN\left(L_k^{\rm r}\right)^2\Big)$.

\section{Simulation Results}\label{Sec:Simulation}
This section provides numerical examples to validate the effectiveness of our proposed algorithm. It is assumed that the BS is located at $[0,0]^T$ and the users are randomly dispersed within a disk centered at $[60,0]^T$ in meters (m) with a radius of $20$ m. The distance between user $k$ and the BS is denoted by $d_k$. We adopt the geometry channel model, where the numbers of transmit and receive channel paths for each user are identical, i.e., $L_k^{\rm t} = L_k^{\rm r} \triangleq L$, $\forall k \in\mathcal K$. Under this condition, the path-response matrix for each user is diagonal, i.e., $\mathbf \Sigma_k = {\rm diag}\{\sigma_{k,1},\cdots,\sigma_{k,L}\}$ with $\sigma_{k,\ell}$ satisfying $\sigma_{k,\ell} \sim \mathcal {CN}\left(0,\frac{c_k^2}{L}\right)$, $\ell = 1,\cdots,L$ \cite{2023_Lipeng_uplink}. Moreover, $c_k^2 = C_0d_k^{-\alpha}$, where $C_0 = -40$ dB is the expected average channel power gain at the reference distance of $1$ m and $\alpha = 2.8$ denotes the path-loss exponent. The elevation and azimuth AoDs/AoAs of the channel paths for each user are assumed to be independent and identically distributed variables, following a uniform distribution over $\left[-\frac{\pi}{2}, \frac{\pi}{2}\right]$. The moving regions for the transmit/receive MAs are set as $\mathcal C^{\rm t} = \mathcal C_k^{\rm r} = \left[ - \frac{A}{2}, \frac{A}{2}\right]\times \left[ - \frac{A}{2}, \frac{A}{2}\right]$, $\forall k\in\mathcal K$. Besides, we set $M = 4$, $D = \frac{\lambda}{2}$, $\sigma_k^2 = -80$ dBm, and $\gamma_k = 1$ for all $k\in\mathcal K$ (i.e., considering the max-min SNR/SINR). The values of  $P_{\max}$, $A$, $L$, $K$ and $N$ will be specified in the following simulations. 

\begin{figure}[!t]
	\hspace{-1mm}
	\subfigure[]{\label{fig:single_vs_iter_1}
		\includegraphics[scale=0.6]{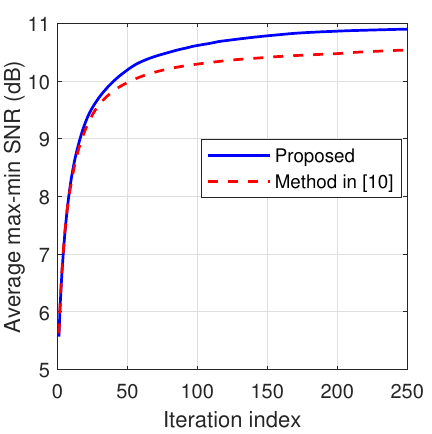}}
	\hspace{-3.5mm}
	\subfigure[]{\label{fig:single_vs_iter_2}
		\includegraphics[scale=0.6]{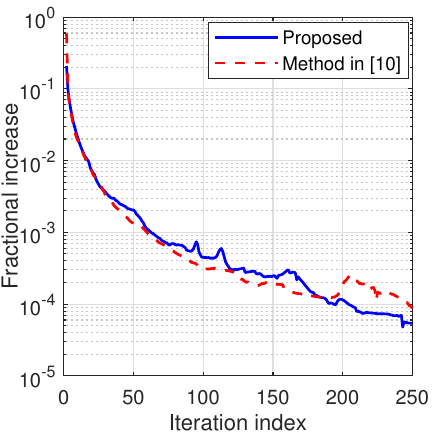}}
	\caption{Convergence behaviors of the proposed Algorithm \ref{Alg1} and the existing method in \cite{2023_Wenyan_MIMO}.
	(a) Average max-min SNR versus the iteration index; (b) The corresponding fractional increase of the max-min SNR versus the iteration index.}
	\label{fig:single_vs_iter}
\end{figure}

\subsection{Single-Group Scenario}
For the single-group scenario, we first illustrate the convergence performance of our proposed Algorithm \ref{Alg1} and compare it with the method based on two-step approximations in \cite{2023_Wenyan_MIMO}, as shown in Fig. \ref{fig:single_vs_iter}. Here, we set $P_{\max} = 15$ dBm, $L = 10$, $A = 3\lambda$, and $K = 6$. In Fig. \ref{fig:single_vs_iter_1}, we observe that the max-min SNR obtained by both algorithms increases with the iteration index, showing particularly rapid growth during the early iterations. Meanwhile, Fig. \ref{fig:single_vs_iter_2} demonstrates a general decreasing trend in the fractional increase of the obtained max-min SNR as the iteration index progresses. Combining Figs. \ref{fig:single_vs_iter_1} and \ref{fig:single_vs_iter_2}, it can be seen that Algorithm \ref{Alg1}, given the threshold $\epsilon$ defined therein, is guaranteed to terminate after a finite number of iterations. 
If we set $\epsilon = 10^{-4}$, Algorithm \ref{Alg1} converges earlier than the method in \cite{2023_Wenyan_MIMO}, yet achieves a superior solution with a 3.4\% performance improvement. This is expected since Algorithm \ref{Alg1} makes only one approximation while the method in \cite{2023_Wenyan_MIMO} utilizes two. 

Next, we compare the achievable max-min SNR of our proposed Algorithm \ref{Alg1} with those of the following four benchmark schemes: 1) \textbf{Receive MA}: the BS is equipped with a uniform linear array (ULA) comprising $M$ FPAs, spaced by $\frac{\lambda}{2}$, while each user employs an MA; 2) \textbf{Transmit MA}: the BS is equipped with $M$ MAs, while the antenna at user $k$ remains fixed at the reference point $\bm o_k^{\rm r} = [0,0]^T$, $\forall k\in\mathcal K$; 3) \textbf{FPA}: the $M$ antennas at the BS and the single antenna at each user are all FPAs; 4) \textbf{Random position}: in each channel realization, we generate $100$ samples of $\left\lbrace\{\bm t_m\}, \{\bm r_k\}\right\rbrace$ randomly and independently. For each position sample, $\left\lbrace\{\bm t_m\}, \{\bm r_k\}\right\rbrace$ satisfy constraints \eqref{P1_cons:d}-\eqref{P1_cons:f} and the transmit beamforming is optimized under given $\left\lbrace\{\bm t_m\}, \{\bm r_k\}\right\rbrace$. We select the best-performing solution among these $100$ samples as the output of this scheme. 

\begin{figure}[!t]
	\centering
	\includegraphics[scale=0.7]{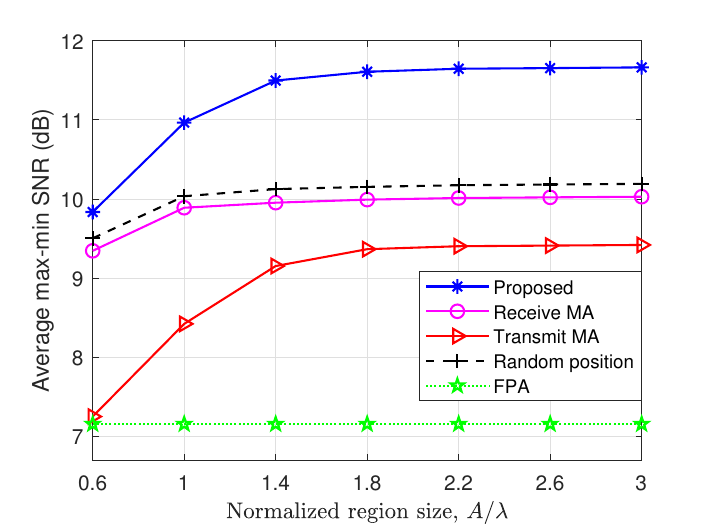}
	\caption{Average max-min SNR versus the normalized region size.} 
	 \label{fig:single_vs_region}
\end{figure}

In Fig. \ref{fig:single_vs_region}, we plot the max-min SNR obtained by different schemes versus the normalized region size $A/{\lambda}$ when $P_{\max} = 15$ dBm, $L = 5$, and $K = 3$. Firstly, it is observed that as $A/{\lambda}$ increases from a small value, the achieved max-min SNR of all the schemes, except for the FPA scheme, exhibits a monotonic growth. This is attributed to the enhanced flexibility in designing the MA positions within the enlarged regions, leading to a potential performance enhancement. However, beyond a certain threshold of $A/{\lambda}$, further increases do not bring performance improvements. This suggests that the maximum achievable max-min SNR is practically attainable within finite transmit and/or receive regions. Secondly, it appears that the max-min SNR achieved by the proposed and transmit MA schemes becomes saturated later than that of the receive MA scheme. In other words, compared to each receive MA, the transmit MAs require a larger moving region to achieve their potential maximum performance. This is because each receive MA has its own moving region, while the $M$ transmit MAs share a single moving region and are constrained by the minimum distance requirement. Thirdly, the three schemes employing transmit and/or receive MAs demonstrate significantly better performance than the FPA scheme. This is anticipated as the MAs can be strategically placed to improve the channel conditions of the users, especially those with relatively poor channel conditions, consequently enhancing the max-min SNR performance. 
Finally, our proposed algorithm consistently achieves the highest max-min SNR since it exploits the most spatial degrees of freedom (DoFs) to enhance the system performance. Take the case with $A/\lambda = 3$ as an example, the proposed algorithm demonstrates approximately 14.4\%, 16.3\%, 23.8\%, and 62.9\% performance improvement over the random position, receive MA, transmit MA, and FPA schemes, respectively. 

\begin{figure}[!t]
	\centering
	\includegraphics[scale=0.7]{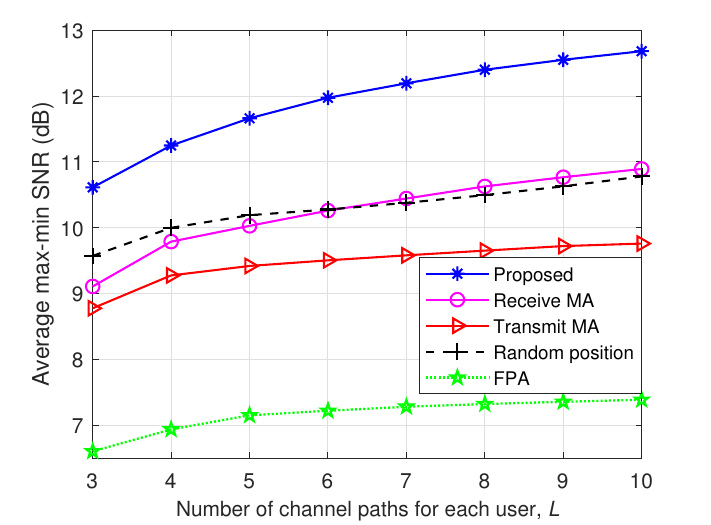}
	\caption{Average max-min SNR versus the number of paths for each user.}
	\label{fig:single_vs_L}
	\vspace{-2mm}
\end{figure}

\begin{figure}[!t]
	\centering
	\includegraphics[scale=0.7]{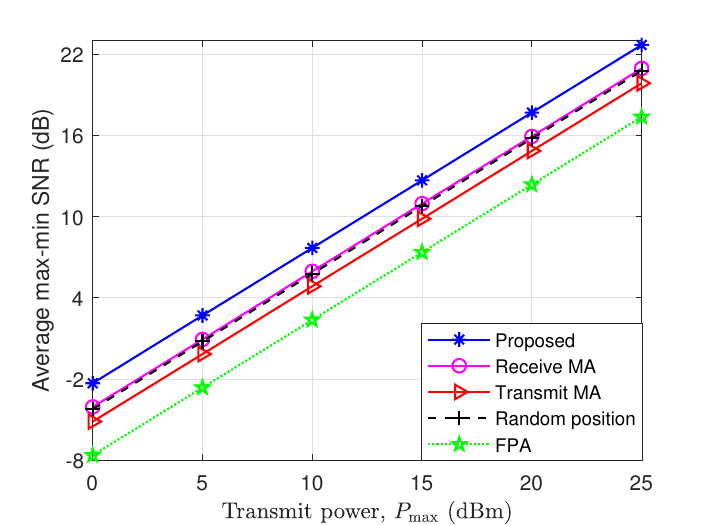}
	\caption{Average max-min SNR versus the maximum transmit power at the BS.\looseness=-1} 
	\label{fig:single_vs_Pmax}
	\vspace{-2mm}
\end{figure}

Fig. \ref{fig:single_vs_L} depicts the max-min SNR versus the number of channel paths $L$ for each user when $P_{\max} = 15$ dBm, $A = 3\lambda$, and $K = 3$. It is first observed that with the increase in $L$, all the schemes, including the FPA scheme, exhibit an enhancement in max-min SNR. This improvement is ascribed to the heightened spatial diversity with increased $L$. Additionally, the MAs can augment spatial diversity further by adjusting their positions within the given moving regions. This is the reason why the max-min SNR experiences a more pronounced increase with $L$ for the four schemes employing MAs compared to the FPA scheme. We also note that the performance of the receive MA scheme surpasses that of the random position scheme when $L$ becomes larger. This is because the receive MA scheme can more effectively capitalize on the enhanced spatial diversity resulting from the larger $L$ by properly placing the receive MAs. As a consequence, it achieves greater performance gains compared to the random position scheme, where the MAs are randomly placed. 

In Fig. \ref{fig:single_vs_Pmax}, we illustrate the max-min SNR of the proposed and benchmark schemes versus the maximum transmit power at the BS when $L = 10$, $A = 3\lambda$, and $K = 3$. As evident, the proposed algorithm can achieve a certain level of performance with reduced transmit power compared to other schemes. For instance, to attain a max-min SNR of $10$ dB, the proposed algorithm requires only about $12$ dBm transmit power, whereas the receive MA, random position, transmit MA, and FPA schemes need about $14$, $14$, $15$, and $17.5$ dBm, respectively. 
It is also noteworthy that the achieved max-min SNR of all the schemes linearly increases with a slope of about $1$ as $P_{\max}$ increases. This is because all the users experience an interference-free stream in the single-group multicast scenario. However, things can be different in the multi-group multicast scenario, which will be clarified in the next subsection. 

\begin{figure}[!t]
	\centering
	\includegraphics[scale=0.7]{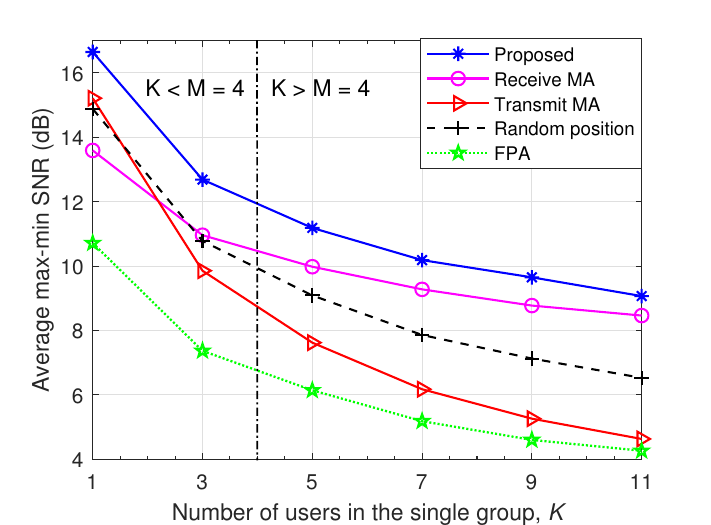}
	\caption{Average max-min SNR versus the number of users in the single group.} 
	\label{fig:single_vs_K}
\end{figure}

\begin{figure}[!ht]
	\centering
	\subfigure[$K = 4$, $N = 2$, and $\left|\mathcal G_1\right| = \left|\mathcal G_2\right|= 2$.]{\label{fig:multi_vs_Pmax_1}
		\includegraphics[scale=0.7]{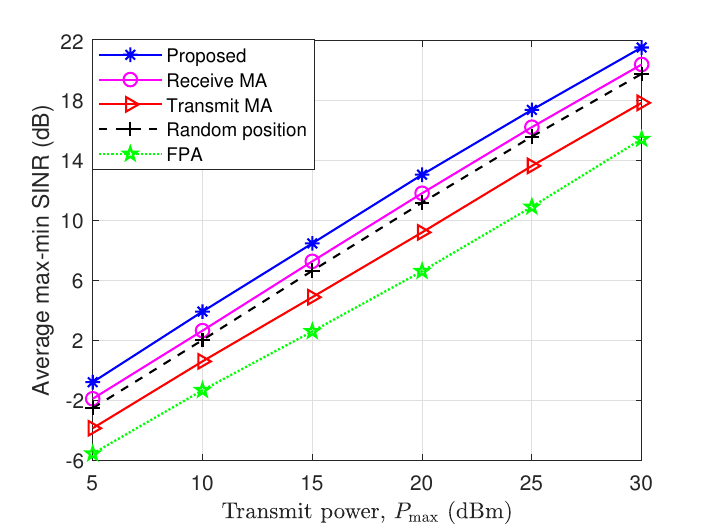}}\vspace{-1mm}
	\subfigure[$K = 9$, $N = 3$, and $\left|\mathcal G_1\right| = \left|\mathcal G_2\right|= \left|\mathcal G_3\right| = 3$.]{\label{fig:multi_vs_Pmax_2}
		\includegraphics[scale=0.7]{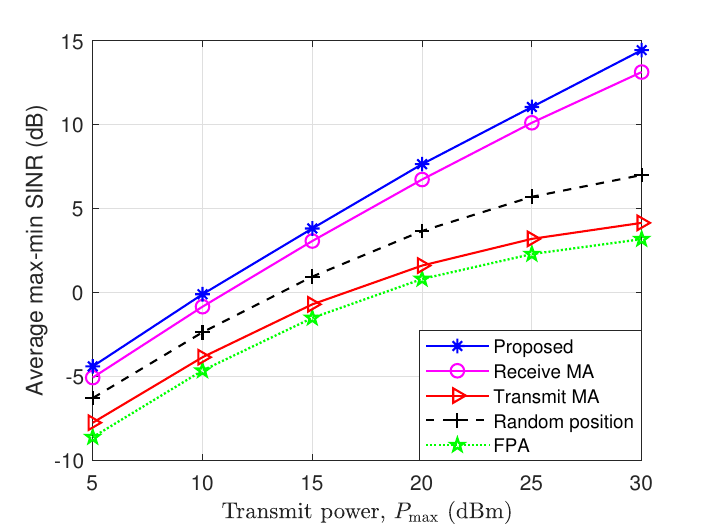}}
	\caption{Average max-min SINR versus the maximum transmit power at the BS. \looseness=-1}
	\label{fig:multi_vs_Pmax}
	\vspace{-1mm}
\end{figure}

Fig. \ref{fig:single_vs_K} investigates the impact of the number of users $K$ on the performance of different schemes when $P_{\max} = 15$ dBm, $L = 10$, and $A = 4\lambda$. Interestingly, it is observed that when $M$ is slightly greater than $K$ (i.e., when $M = 4$ and $K = 3$), the receive MA scheme achieves a higher max-min SINR than the transmit MA scheme, despite the receive MAs being at a numerical disadvantage compared to the transmit MAs. This is primarily because, in the receive MA scheme, all users can achieve their individual maximum SNR simultaneously by adjusting the receive MAs to positions that yield higher channel gains and/or increased correlation among user channel vectors within the same group. In contrast, in the transmit MA scheme, the users are always competing for resources, and the positioning of the transmit MAs requires consideration of the trade-offs among the channel conditions of all users. As such, the receive MA scheme outperforms the transmit MA scheme, even though the transmit MAs are slightly outnumbered. Furthermore, with the increase of $K$, the gap between the curves representing the receive MA and transmit MA schemes widens. This happens because the number of transmit MAs remains constant, while the number of receive MAs increases with $K$. As $K$ increases, it becomes increasingly difficult for the transmit MAs to maintain performance fairness among different users. Nevertheless, it remains feasible for all users to achieve their individual maximum SNR with the aid of receive MAs. As the effect of transmit MAs diminishes, the performance gain of the proposed algorithm compared to the receive MA scheme decreases from 22.5\% at $K=1$ to 7.2\% at $K=11$. 

\subsection{Multi-Group Scenario}
This subsection considers the multi-group scenario and compares the performance of our proposed algorithm with the four benchmark schemes defined in the previous subsection. 

In Fig. \ref{fig:multi_vs_Pmax}, we plot the max-min SINR versus the transmit power at the BS when $L = 10$ and $A = 4\lambda$. Two setups are considered: one with $K = 4$, $N = 2$, and $\left|\mathcal{G}_1\right| = \left|\mathcal{G}_2\right| = 2$ (Fig. \ref{fig:multi_vs_Pmax_1}), and the other with $K = 9$, $N = 3$, and $\left|\mathcal{G}_1\right| = \left|\mathcal{G}_2\right| = \left|\mathcal{G}_3\right| = 3$ (Fig. \ref{fig:multi_vs_Pmax_2}). From Fig. \ref{fig:multi_vs_Pmax_1}, it is observed that as $P_{\max}$ increases, all the curves, including the one representing the FPA scheme, do not show a deceleration in growth rate or signs of saturation. This is in line with the result in \cite{2017_Hamdi_multicast}, which indicates that if the number of transmit antennas $M$ satisfies the condition: $M \geq 1 + K - \min_{n\in\mathcal N}\left|\mathcal G_n\right|$, classical beamforming enables each beam to be placed in the null space of all its unintended groups. As such, each multicast group can receive an interference-free stream. However, the simulation setup of Fig. \ref{fig:multi_vs_Pmax_2} violates the condition that $M \geq 1 + K - \min_{n\in\mathcal N}\left|\mathcal G_n\right|$. Therefore, as the inter-group interference increases with $P_{\max}$, the max-min SINR achieved by the FPA scheme tends to saturate. In contrast, the max-min SINR of the proposed algorithm exhibits an almost linear increase with the growth of $P_{\max}$ in both Fig. \ref{fig:multi_vs_Pmax_1} and Fig. \ref{fig:multi_vs_Pmax_2}, significantly surpassing that of the FPA scheme. This superiority can be attributed to the advantages of MAs, which not only enhance the channel gain but also mitigate inter-group interference more effectively compared to the FPAs. Besides, we note that the receive MA scheme performs much better than the transmit MA scheme, with the performance gain being more pronounced in Fig. \ref{fig:multi_vs_Pmax_2} compared to that in Fig. \ref{fig:multi_vs_Pmax_1}.  This suggests that when $K \geq M$, receive MAs are more effective in improving max-min fairness than transmit MAs, and the larger the gap between $K$ and $M$, the more obvious it is. This aligns with the conclusion drawn from Fig. \ref{fig:single_vs_K}. \looseness=-1

\begin{figure}[!t]
	\centering
	\includegraphics[scale=0.7]{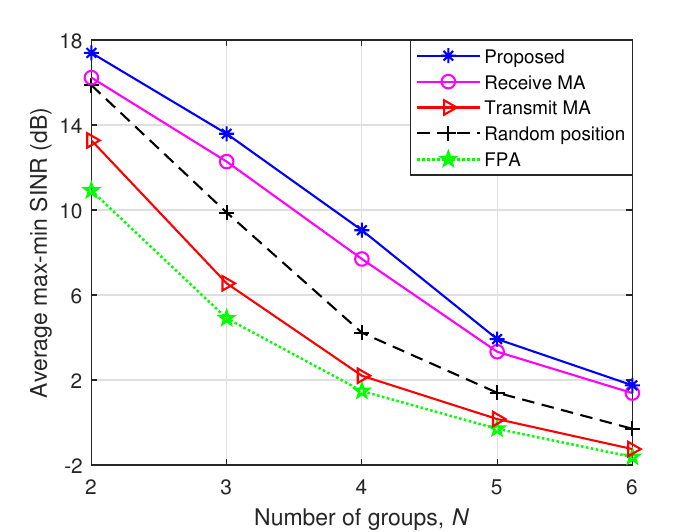}
	\caption{Average max-min SINR versus the number of groups.} \label{fig:multi_vs_Gnum}
	\vspace{-1mm}
\end{figure} 

Fig. \ref{fig:multi_vs_Gnum} illustrates the max-min SINR achieved by different schemes versus the number of groups when $P_{\max} = 25$ dBm, $L =10$, and $A = 4\lambda$. Here, we assume that each group consists of $2$ users, i.e., $\left|\mathcal G_n\right| = 2$, $\forall n\in\mathcal N$. It is observed that there is a significant decrease in the achieved max-min SINR with increasing $N$ for all schemes. This is intuitive since a greater number of groups makes it more difficult to mitigate inter-group interference. Particularly, based on the DoF analyses in \cite{2017_Hamdi_multicast}, the MMF-DoF achieved by classical beamforming with FPAs decreases from $1$ to $0$ when $N$ increases from $2$ to $3$. Thus, the polyline representing the FPA scheme displays the steepest slope during the transition from $N = 2$ to $N = 3$, compared to other variations in $N$. By contrast, the steepest slope for the proposed algorithm occurs during the transition from $N = 4$ to $N = 5$. The reason lies in the proposed algorithm's ability to leverage the diversity gain and the interference mitigation gain in the spatial domain by adjusting the positions of both transmit and receive MAs. This feature makes it very likely for the proposed algorithm to have positive MMF-DoF values even when $N=3$ and $N=4$. It is also noteworthy that in overloaded scenarios (i.e., $N > M = 4$), the performance gain from moving the transmit MAs is minor since the limitation imposed by the number of transmit antennas cannot be overcome. In contrast, each receive MA can still be repositioned to enhance the SINR of its corresponding user, thereby improving the system's max-min SINR.

\begin{figure}[!t]
	\centering
	\includegraphics[scale=0.7]{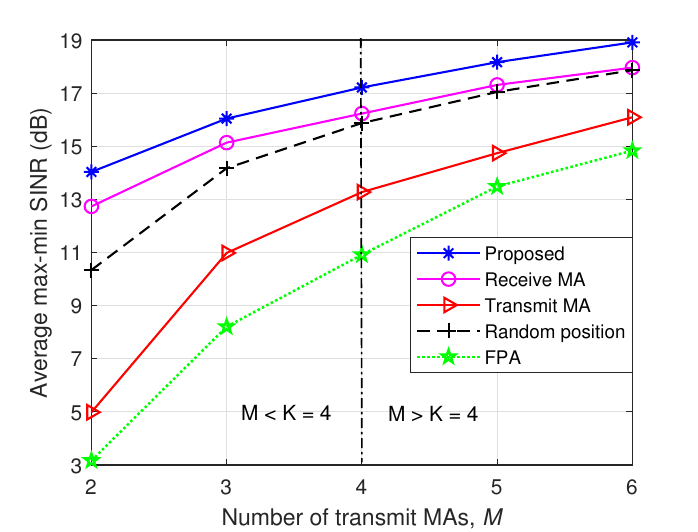}
	\caption{Average max-min SINR versus the number of transmit MAs at the BS.} \label{fig:multi_vs_M}
	\vspace{-2mm}
\end{figure}

The effect of the number of transmit MAs, $M$, on the performance of different schemes is shown in Fig. \ref{fig:multi_vs_M}. Other parameters are set as: $P_{\max} = 25$ dBm, $L = 10$, $A = 4\lambda$, $K = 4$, $N = 2$, and $\left|\mathcal G_1\right| = \left|\mathcal G_2\right| = 2$. It is shown that the increase in $M$ results in an augmentation of the max-min SINR achieved by all the schemes. This is expected as the additional antennas introduce more DoFs, allowing the BS to form sharp beams directed toward the desired users to improve the desired signal power with less co-channel interference. Moreover, the proposed algorithm outperforms the others under any value of $M$. In particular, when $M = 2$, the max-min SINR of the proposed algorithm increases by 181.7\% and 343.9\% compared to the transmit MA and FPA schemes, respectively. Also, the proposed algorithm requires fewer transmit antennas than other schemes to achieve a certain level of max-min SINR performance due to its higher flexibility in MA positioning. Lastly, when $M$ slightly exceeds $K$, the receive MA scheme still performs better than the transmit MA scheme, which is in consistent with the observation from Fig. \ref{fig:single_vs_K}. 

\section{Conclusion}\label{Sec:Conclusion}
In this paper, we investigated a novel MA-enhanced multicast communication system. Aiming to maximize the minimum weighted SINR among all the users, we formulated a design problem, where the position of each transmit/receive MA and the transmit precoders at the BS were jointly optimized. Despite the non-convex nature of the considered problem with highly coupled optimization variables, we proposed an efficient AO-based algorithm initially customized for the single-group scenario, which was then extended to the general multi-group scenario. Our proposed algorithm was validated through simulations, demonstrating its effectiveness and offering useful engineering insights. Firstly, employing both transmit and receive MAs results in max-min SNR/SINR improvements of up to 22.5\%, 181.7\%, and 343.9\% over benchmarks that use only receive MAs, only transmit MAs, and both transmit and receive FPAs, respectively. Secondly, compared to benchmark schemes, the proposed algorithm achieves a significant reduction in the necessary transmit power or number of antennas to attain a target level of max-min SNR/SINR performance. Thirdly, in overloaded scenarios, using only transmit MAs offers minimal max-min SINR improvement over using both transmit and receive FPAs. Conversely, using only receive MAs still provides notable performance gains, as exemplified by an improvement in max-min SINR from $-0.29$ dB to $3.33$ dB. Fourthly, under the assumption that the transmit region and each receive region have identical sizes, employing only receive MAs outperforms employing only transmit MAs in terms of max-min SNR/SINR when $K \geq M$. And, this remains true even when $M$ is slightly greater than $K$. Finally, our simulations revealed that some theoretical results derived for traditional FPA-enabled multicast systems do not apply to MA-enabled multicast systems. Developing suitable theoretical results for these new systems is a challenging issue that necessitates further study.

\appendix[Construction of $\psi_{k,m}$]
First, we derive the Hessian matrix of $u_k(\boldsymbol t_m)$, $\nabla^2 u_k(\boldsymbol t_m)$, as
\begin{align}
	\nabla^2 u_k(\boldsymbol t_m) = \begin{bmatrix} \frac{\partial u_k(\boldsymbol t_m)}{\partial x_m^{\rm t}\partial x_m^{\rm t}} & \frac{\partial u_k(\boldsymbol t_m)}{\partial x_m^{\rm t}\partial y_m^{\rm t}} \\ \frac{\partial u_k(\boldsymbol t_m)}{\partial y_m^{\rm t}\partial x_m^{\rm t}} & \frac{\partial u_k(\boldsymbol t_m)}{\partial y_m^{\rm t}\partial y_m^{\rm t}} \end{bmatrix},
\end{align} where the expressions of the matrix's elements are provided in \eqref{eq:second-order} at the top of the subsequent page. 
\begin{figure*}[!ht]
	\begin{subequations}\label{eq:second-order}
		\begin{align}
			\frac{\partial u_k(\boldsymbol t_m)}{\partial x_m^{\rm t}\partial x_m^{\rm t}} = & -\frac{8\pi^2}{\lambda^2}\sum_{i = 1}^{L_k^{\rm t}-1}\sum_{j = i+1}^{L_k^{\rm t}}\left|w_m \right|^2\left|\left[\bm B_k\right]_{i,j}\right|\left(- \cos\theta_{k,i}^{\rm t}\sin\phi_{k,i}^{\rm t} + \cos\theta_{k,j}^{\rm t}\sin\phi_{k,j}^{\rm t}\right)^2\cos\left(\beta_{i,j,k}(\bm t_m)\right) \nonumber\\
			& -\frac{8\pi^2}{\lambda^2}\sum_{i = 1}^{L_k^{\rm t}}\left|w_m\right|\left|\Lambda_{k,m}\right|\left|\left[\bm b_k\right]_i\right|\cos^2\theta_{k,i}^{\rm t}\sin^2\phi_{k,i}^{\rm t}\cos\left(\iota_{k,i}(\bm t_m)\right), \\
			\frac{\partial u_k(\boldsymbol t_m)}{\partial x_m^{\rm t}\partial y_m^{\rm t}} = & \frac{\partial u_k(\boldsymbol t_m)}{\partial y_m^{\rm t}\partial x_m^{\rm t}} =  -\frac{8\pi^2}{\lambda^2}\sum_{i = 1}^{L_k^{\rm t}-1}\sum_{j = i+1}^{L_k^{\rm t}}\left|w_m \right|^2\left|\left[\bm B_k\right]_{i,j}\right|\left(-\cos\theta_{k,i}^{\rm t}\sin\phi_{k,i}^{\rm t} + \cos\theta_{k,j}^{\rm t}\sin\phi_{k,j}^{\rm t}\right)\left(-\sin\theta_{k,i}^{\rm t} + \sin\theta_{k,j}^{\rm t}\right)\nonumber\\
			& \hspace{1.8cm}\times \cos\left(\beta_{i,j,k}(\bm t_m)\right) -\frac{8\pi^2}{\lambda^2}\sum_{i = 1}^{L_k^{\rm t}}\left|w_m\right|\left|\Lambda_{k,m}\right|\left|\left[\bm b_k\right]_i\right|\cos\theta_{k,i}^{\rm t}\sin\phi_{k,i}^{\rm t}\sin\theta_{k,i}^{\rm t}\cos\left(\iota_{k,i}(\bm t_m)\right), \\
			\frac{\partial u_k(\boldsymbol t_m)}{\partial y_m^{\rm t}\partial y_m^{\rm t}} = & -\frac{8\pi^2}{\lambda^2}\sum_{i = 1}^{L_k^{\rm t}-1}\sum_{j = i+1}^{L_k^{\rm t}}\left|w_m \right|^2\left|\left[\bm B_k\right]_{i,j}\right|\left(-\sin\theta_{k,i}^{\rm t} + \sin\theta_{k,j}^{\rm t}\right)^2\cos\left(\beta_{i,j,k}(\bm t_m)\right) \nonumber\\
			& -\frac{8\pi^2}{\lambda^2}\sum_{i = 1}^{L_k^{\rm t}}\left|w_m\right|\left|\Lambda_{k,m}\right|\left|\left[\bm b_k\right]_i\right|\sin^2\theta_{k,i}^{\rm t}\cos\left(\iota_{k,i}(\bm t_m)\right).
		\end{align}	\hrulefill
	\end{subequations}
\end{figure*}
Second, we have 
\begin{align}\label{ineq:psi}
	&\left\|\nabla^2 u_k(\boldsymbol t_m)\right\|_2 \leq \left\|\nabla^2 u_k(\boldsymbol t_m)\right\|_F \nonumber\\
	& = \Bigg[\left(\frac{\partial u_k(\boldsymbol t_m)}{\partial x_m^{\rm t}\partial x_m^{\rm t}}\right)^2 + \left( \frac{\partial u_k(\boldsymbol t_m)}{\partial x_m^{\rm t}\partial y_m^{\rm t}}\right)^2 + \left( \frac{\partial u_k(\boldsymbol t_m)}{\partial y_m^{\rm t}\partial x_m^{\rm t}}\right)^2 \nonumber\\
	& \hspace{4mm} + \left(\frac{\partial u_k(\boldsymbol t_m)}{\partial y_m^{\rm t}\partial y_m^{\rm t}}\right)^2\Bigg]^{\frac{1}{2}}. 
\end{align}
Note that the terms $\frac{\partial u_k(\boldsymbol t_m)}{\partial x_m^{\rm t}\partial x_m^{\rm t}}$, $\frac{\partial u_k(\boldsymbol t_m)}{\partial x_m^{\rm t}\partial y_m^{\rm t}}$, $\frac{\partial u_k(\boldsymbol t_m)}{\partial x_m^{\rm t}\partial y_m^{\rm t}}$, and $\frac{\partial u_k(\boldsymbol t_m)}{\partial y_m^{\rm t}\partial y_m^{\rm t}}$ all comprise unknown $\bm t_m$-related components. By setting $\cos\left(\beta_{i,j,k}(\bm t_m)\right) = \cos\left(\iota_{k,i}(\bm t_m)\right) = 1$, we can obtain an upper bound of $\left[\left(\frac{\partial u_k(\boldsymbol t_m)}{\partial x_m^{\rm t}\partial x_m^{\rm t}}\right)^2 + \left( \frac{\partial u_k(\boldsymbol t_m)}{\partial x_m^{\rm t}\partial y_m^{\rm t}}\right)^2 + \left( \frac{\partial u_k(\boldsymbol t_m)}{\partial y_m^{\rm t}\partial x_m^{\rm t}}\right)^2 + \left(\frac{\partial u_k(\boldsymbol t_m)}{\partial y_m^{\rm t}\partial y_m^{\rm t}}\right)^2\right] ^{\frac{1}{2}}$, denoted by $\bar\psi_{k,m}$. Since $\bar\psi_{k,m} \geq \left\|\nabla^2 u_k(\boldsymbol t_m)\right\|_2$ and $\left\|\nabla^2 u_k(\boldsymbol t_m)\right\|_2\mathbf I \succeq \nabla^2 u_k(\boldsymbol t_m)$, we have $\bar\psi_{k,m}\mathbf I \succeq \nabla^2 u_k(\boldsymbol t_m)$. Thus, we can choose $\psi_{k,m}$ to be equal to $\bar\psi_{k,m}$. 

\bibliographystyle{IEEEtran}
\bibliography{ref}

\begin{IEEEbiography}[{\includegraphics[width=1in,height=1.25in,clip,keepaspectratio]{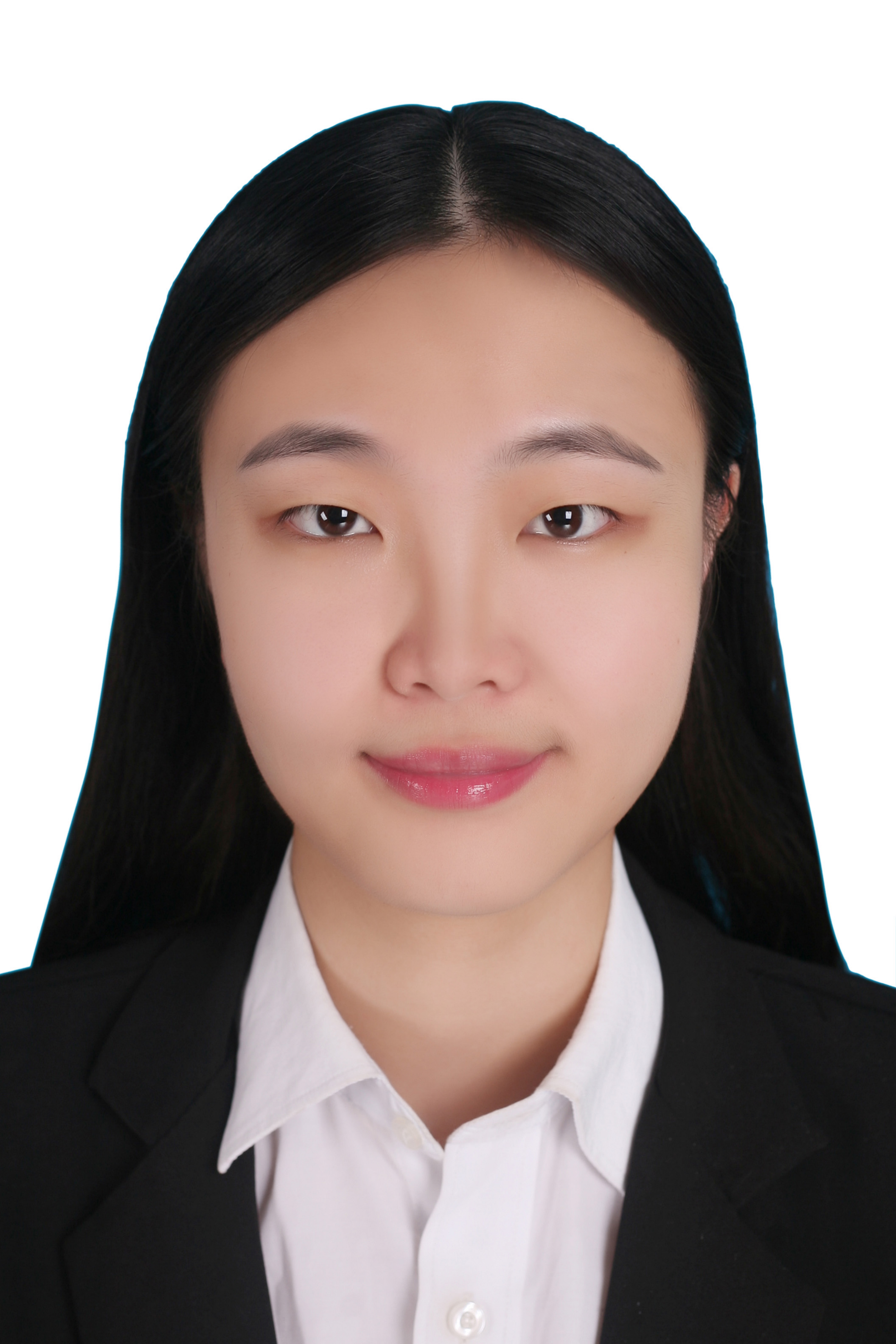}}]{Ying Gao} received the B.Eng. degree from Nanjing University of Science and Technology (NJUST), Nanjing, China, in 2016, and the Ph.D. degree from Shanghai Institute of Microsystem and Information Technology (SIMIT), Chinese Academy of Sciences (CAS), Shanghai, China, in 2021. From August 2021 to November 2023, she held the position of Post-Doctoral Research Fellow with the State Key Laboratory of Internet of Things for Smart City, University of Macau (UM). She is currently a Post-Doctoral Research Fellow with the Department of Electronic Engineering, Shanghai Jiao Tong University (SJTU). Her current research interests include intelligent reflecting surface (IRS) assisted communications, movable antenna (MA) enhanced communications, unmanned aerial vehicle (UAV) enabled communications, physical layer security, and optimization theory. 
\end{IEEEbiography}

\begin{IEEEbiography}[{\includegraphics[width=1in,height=1.25in,clip,keepaspectratio]{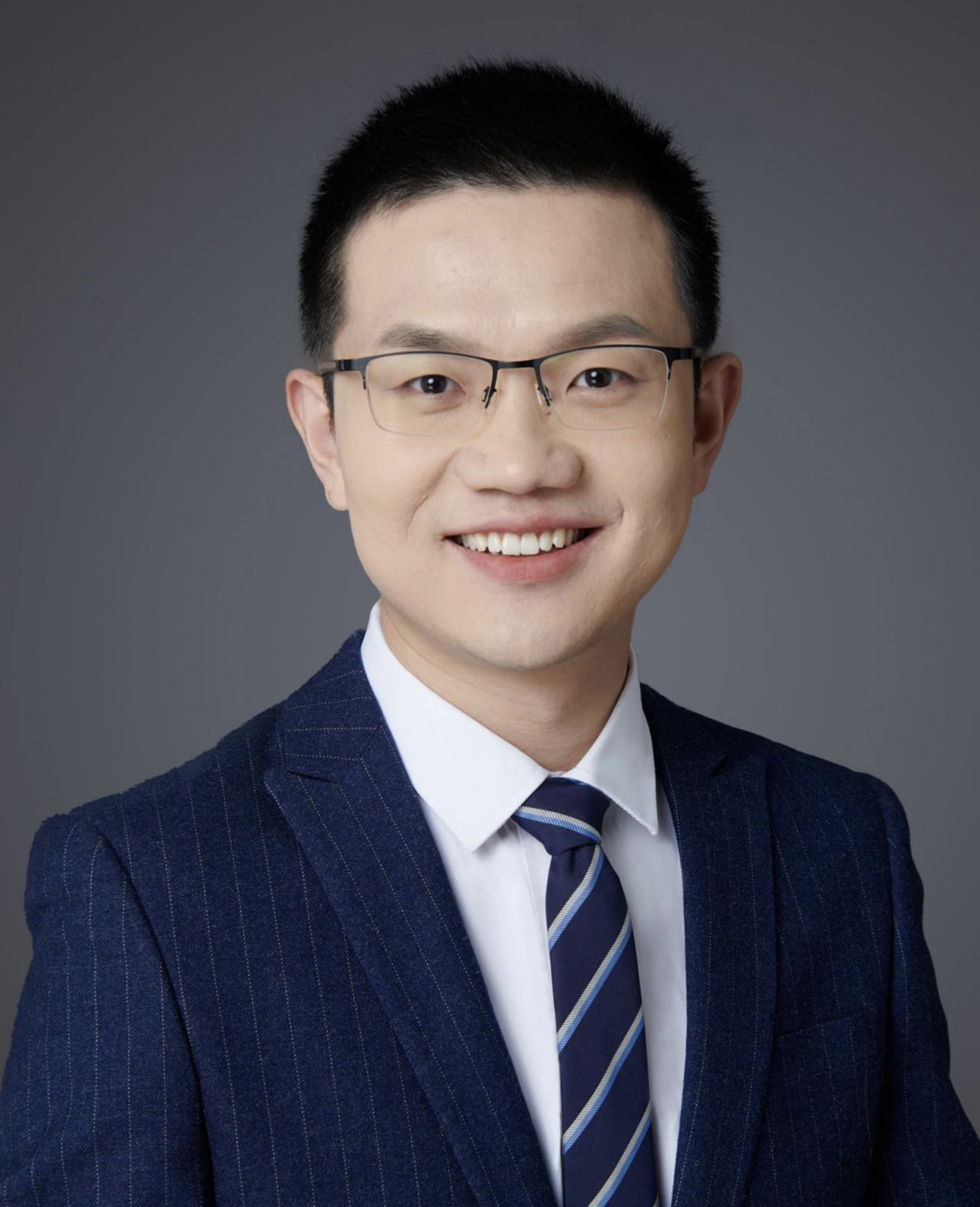}}]{Qingqing Wu} (S’13-M’16-SM’21)  is an Associate Professor with Shanghai Jiao Tong University. His current research interest includes intelligent reflecting surface (IRS), unmanned aerial vehicle (UAV) communications, and MIMO transceiver design. He has coauthored more than 100 IEEE journal papers with 40+ ESI highly cited papers, which have received more than 32,000 Google citations. He has been listed as the Clarivate ESI Highly Cited Researcher since 2021.

He was the recipient of the IEEE Communications Society Fred Ellersick Prize, IEEE  Best Tutorial Paper Award in 2023,  and IEEE WCSP Best Paper Award in 2015. He serves as an Associate/Senior/Area Editor for IEEE Transactions on Wireless Communications, IEEE Transactions on Communications, IEEE Communications Letters, IEEE Wireless Communications Letters. He is the Lead Guest Editor for IEEE Journal on Selected Areas in Communications. He is the workshop co-chair for IEEE ICC 2019-2023 and IEEE GLOBECOM 2020.  He is the Founding Chair of  IEEE Communications Society Young Professional committee in Asia Pacific Region and Chair of IEEE VTS Drone Committee. 
\end{IEEEbiography}

\begin{IEEEbiography}[{\includegraphics[width=1in,height=1.25in,clip,keepaspectratio]{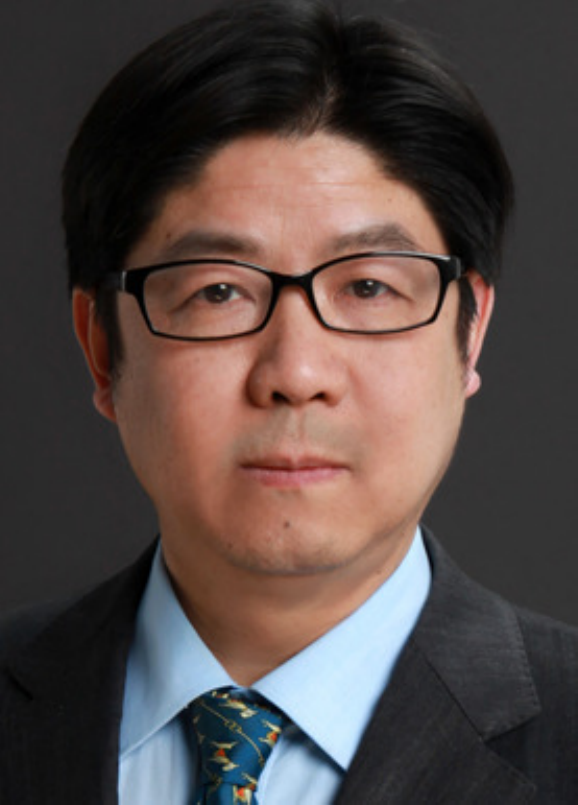}}] {Wen Chen} (M’03–SM’11) received B.S. and M.S. from Wuhan University, China in 1990 and 1993 respectively, and Ph.D. from University of Electro-communications, Japan in 1999. He is now a tenured Professor with the Department of Electronic Engineering, Shanghai Jiao Tong University, China, where he is the director of Broadband Access Network Laboratory. He is a fellow of Chinese Institute of Electronics and the distinguished lecturers of IEEE Communications Society and IEEE Vehicular Technology Society. He is the Shanghai Chapter Chair of IEEE Vehicular Technology Society, a vice president of Shanghai Institute of Electronics, Editors of IEEE Transactions on Wireless Communications, IEEE Transactions on Communications, IEEE Access and IEEE Open Journal of Vehicular Technology. His research interests include multiple access, wireless AI and RIS communications. He has published more than 200 papers in IEEE journals with citations more than10,000 in Google scholar. 
\end{IEEEbiography}	

\end{document}